\documentclass[a4paper,11pt,bibtotoc,liststotoc]{article}

\usepackage{amssymb}
\usepackage{graphicx}
\usepackage{hyperref}
\usepackage{harvard}
\usepackage{amsmath}
\usepackage[english]{babel}
\usepackage{amsfonts}
\usepackage{rotating}
\usepackage{lscape}
\usepackage[official]{eurosym}
\usepackage{longtable}
\usepackage[a4paper,left=2.5cm,right=2.5cm,top=2.5cm,bottom=2.5cm]{geometry}
\usepackage{xspace}
\usepackage{array}
\usepackage{latexsym}
\usepackage{rotating}
\usepackage[toc,page]{appendix}
\usepackage{tkz-graph}
\usetikzlibrary{shapes.geometric}%

\usepackage{xcolor}
\usepackage{booktabs}
\usepackage{fancyref}

\usepackage{amsmath,amssymb}

\usepackage{accents}

\newenvironment{packed_item}{
\begin{itemize}
  \setlength{\itemsep}{1pt}
  \setlength{\parskip}{0pt}
  \setlength{\parsep}{0pt}
}{\end{itemize}}
\usepackage{multirow}
\usepackage{lscape}
\usepackage{color}
\usepackage{soul}
\usepackage{typearea}
\newcommand{\floor}[1]{\left\lfloor #1 \right\rfloor}

\newcommand{\ben}{\begin{enumerate}}
\newcommand{\een}{\end{enumerate}}
\newcommand{\bi}{\begin{itemize}}
\newcommand{\ei}{\end{itemize}}
\newcommand{\bd}{\begin{description}}
\newcommand{\ed}{\end{description}}

\newcommand{\bear}{\begin{eqnarray}}
\newcommand{\eear}{\end{eqnarray}}

\usepackage{scrpage2}
\setheadwidth{textwithmarginpar}
\usepackage{caption}
\usepackage{subfig}
\usepackage{setspace}
\usepackage{xspace}
\usepackage{authblk}

\usepackage{soul}

\begin{document}

\begin{center}

{\LARGE Bounds on direct and indirect effects\\under treatment/mediator endogeneity and outcome attrition
}

{\large \vspace{0.1cm}}

{\Large Martin Huber* and Luk\'{a}\v{s} Laff\'{e}rs**}\smallskip

{\small {*University of Fribourg, Dept.\ of Economics} \\
	**Matej Bel University, Dept. of Mathematics\bigskip }
	
\end{center}

\smallskip

\noindent \textbf{Abstract:} {\small Causal mediation analysis aims at disentangling a treatment effect into an indirect mechanism operating through an intermediate outcome or mediator, as well as the direct effect of the treatment on the outcome of interest.  However, the evaluation of direct and indirect effects is frequently complicated by  non-ignorable selection into the treatment and/or mediator, even after controlling for observables, as well as sample selection/outcome attrition. We propose a method for bounding direct and indirect effects in the presence of such complications using a method that is based on a sequence of linear programming problems. Considering inverse probability weighting by propensity scores, we compute the weights that would yield identification in the absence of complications and perturb them by an entropy parameter reflecting a specific amount of propensity score misspecification to set-identify the effects of interest. We apply our method to data from the National Longitudinal Survey of Youth 1979 to derive bounds on the explained and unexplained components of a gender wage gap decomposition that is likely prone to non-ignorable mediator selection and outcome attrition.}

{\small \smallskip }



{\small \smallskip {\scriptsize 
	Addresses for correspondence: Martin Huber, University of Fribourg, Bd.\ de P\'{e}rolles 90, 1700 Fribourg, Switzerland; martin.huber@unifr.ch. Luk\'{a}\v{s}	Laff\'{e}rs, Department of Mathematics, Matej Bel University, Tajovsk\'{e}ho 40, SK97401 Bansk\'{a} Bystrica, Slovakia. Laff\'{e}rs acknowledges support provided by the  Slovak  Research  and  Development  Agency under the contract No. APVV-17-0329 and VEGA-1/0692/20.
}\thispagestyle{empty}\pagebreak  }

{\small \renewcommand{\thefootnote}{\arabic{footnote}} %
\setcounter{footnote}{0}  \pagebreak \setcounter{footnote}{0} \pagebreak %
\setcounter{page}{1} }

\section{Introduction}

Mediation analysis aims to decompose a treatment effect into an indirect causal mechanism operating through one or several intermediate variables, so-called mediators, as well as the direct effect,including any mechanisms not operating through the mediators of interest. For instance, early childhood interventions might affect labor market or health outcomes later in life through different mechanisms like the formation of cognitive or non-cognitive skills, see for instance  \citeasnoun{HeckmanPintoSavelyev2013} and \citeasnoun{Keeleetal2015}. Furthermore, job seeker counseling may influence employment through assignment to training programs or other mechanisms in the counseling process, see \citeasnoun{HuberLechnerMellace2017}. Even with a randomly assigned treatment, direct and indirect effects are generally not identified by naively controlling for mediators, as this likely introduces selection bias, see \citeasnoun{RoGr92}. While much of the earlier work on mediation analysis assumed linear models and/or neglected selection issues,  see \citeasnoun{Co57}, \citeasnoun{JuKe81}, and \citeasnoun{BaKe86}, more recent contributions discuss more general identification approaches and explicitly consider confounding. See for instance \citeasnoun{RoGr92}, \citeasnoun{Pearl01}, \citeasnoun{Robins2003},  \citeasnoun{PeSiva06}, \citeasnoun{VanderWeele09}, \citeasnoun{ImKeYa10}, \citeasnoun{Hong10}, \citeasnoun{AlbertNelson2011}, \citeasnoun{ImYa2011},  \citeasnoun{TchetgenTchetgenShpitser2011}, and \citeasnoun{VansteelandtBekaertLange2012}.

In most mediation studies, identification relies on a conditionally exogenous treatment and mediator given observed covariates and rules out non-ignorable outcome attrition or sample selection, i.e.\ that outcomes are only observed for a nonrandom subpopulation. This issue occurs for instance in wage regressions, where wages are only observed for the selective subgroup of employed individuals,  see \citeasnoun{He76} and \citeasnoun{Heckman79}. For this reason, \citeasnoun{HuberSolovyeva2017} incorporate outcome attrition into mediation models, assuming conditional treatment and mediator exogeneity and tackling outcome attrition either by observed covariates (missing at random assumption, see e.g.\  \citeasnoun{Ru76b} and \citeasnoun{LittleRubin87})
or by instruments (if attrition is selective in unobservables). In many empirical problems, however, observed covariates might not be rich enough to convincingly control for treatment/mediator endogeneity and attrition bias while instruments that satisfy specific exclusion restrictions w.r.t.\ attrition (see for instance \citeasnoun{Daneve03} and \citeasnoun{Hu11b}) might not be available.

This paper provides a method for deriving bounds on direct and indirect effects when the treatment, the mediator and outcome attrition are likely selective even after controlling for observed covariates. Considering identification based on inverse probability weighting based on a combination of propensity scores, we compute the weights that would yield identification in the absence of complications (as provided in \citeasnoun{HuberSolovyeva2017}) and perturb them by an entropy parameter reflecting misspecification in the various propensity scores. Based on the framing the identification issue as an optimization problem to be solved by linear programming, we set-identify the mean potential outcomes and thus, the direct and indirect effects of interest.

Our contribution is related to further studies that used optimization, and in particular linear programming, to derive bounds on treatment effects under selection problems, see e.g.\ \citeasnoun{BaPe97}, \citeasnoun{manski2007partial},  \citeasnoun{honore2006bounds}, \citeasnoun{molinari2008partial}, \citeasnoun{freyberger2015identification}, \citeasnoun{laffers2017sensitivity}, \citeasnoun{laffers2019bounding}, among many others. Our paper is also related to the literature on sensitivity analysis in mediation analysis.  \citeasnoun{VanderWeele2010}, for instance, provides a general formula for the bias of direct and indirect effects in the presence of an unobserved mediator-outcome confounder. By considering sensible values for differences in conditional mean outcomes across confounder values and for differences in the conditional mean of the confounder across treatment states, researches may investigate the sensitivity of the effects. \citeasnoun{ImKeYa10} propose a sensitivity check for parametric (both linear and nonlinear) mediation models based on specifying the correlation of unobserved terms in the mediation and outcome equations, assuming that the mediator-outcome confounders are not a function of the treatment. In contrast, \citeasnoun{TchetgenTchetgenShpitser2011} suggest a semiparametric procedure that allows for confounders of the mediator-outcome relation which are affected by the treatment based on specifying and calibrating the so-called selection bias function, which is agnostic about the dimension of unobserved confounders. See \citeasnoun{VanderWeeleChiba2014} and \citeasnoun{VansteelandtVanderWeele2012} for further selection bias functions.

As an alternative strategy, \citeasnoun{AlbertNelson2011} suggest considering the correlation of counterfactual values of post-treatment variables as sensitivity parameter. Finally, the paper that is the closest to our approach is \citeasnoun{HongQinYang2018}, which provides a method tailored to weighting estimators under the omission of both pre- and post-treatment confounders. The idea is that such confounders create a discrepancy between the correct weight an observation should obtain and the one actually used. The resulting bias can be represented by the covariance between the weight discrepancy and the outcome conditional on the treatment, which serves as base for conducting sensitivity analyses. Our approach is different to \citeasnoun{HongQinYang2018} in that   we represent the discrepancy between the correct and observed weights using entropy parameters and, instead of deriving analytical formulas, rely on an optimization routine to obtain bounds on direct and indirect effects. This allows for a separate relaxation of the three main identification assumptions and thus may lead to a better understanding of the non-robustness of the results to violations of the various identification assumptions. We also note that the econometric setup of \citeasnoun{HuberSolovyeva2017} underlying our analysis invokes a different set of assumptions than \citeasnoun{HongQinYang2018} or any of the other previous methods, such that our approach permits investigating sensitivity also w.r.t.\ outcome attrition.\footnote{The studies mentioned and our own investigate the sensitivity of direct and indirect effects to prespecified deviations from the identifying assumptions. Alternatively, one may derive worst case bounds, which are based on the possibly most extreme forms of violations of specific assumptions, which typically implies a rather wide range of admissible effect values. See for instance \citeasnoun{Kaufmanetal05}, \citeasnoun{Caietal08},  \citeasnoun{Sj09}, and \citeasnoun{FlFl10}.}

We apply our method to data from the  National Longitudinal Survey of Youth 1979, a panel study of young individuals in the U.S.\ aged 14 to 22 years in 1979. The specific sample considered has previously been analyzed by \citeasnoun{HuberSolovyeva2019} to decompose the gender gap in wages reported in the year 2000 into an indirect (or explained) component due to differences in mediators like education and occupation, as well as a direct (or unexplained) gender difference in wages not attributable to the observed mediators. While \citeasnoun{HuberSolovyeva2019} investigated the sensitivity of point estimation of explained and unexplained component under different identifying assumptions, our approach permits easing any of the conditional exogeneity assumptions on gender, the mediators, and selection into employment (as wages are only observed for working individuals) to derive bound son the parameters of interest. We find that the omission of confounders of the treatment and the mediators would potentially have the largest impact on the significance of the results. More specifically, the omission of a confounder that has the same predictive power as the first or second most important mediator entering the treatment propensity score would render all the effects insignificant. The results also show that in some specifications the choice of the link function in the estimation of probabilistic weights matter.

The remainder of this paper is organized as follows. Section \ref{param} introduces the variables as well as the direct and indirect effects of interest. Section  \ref{sensitivityan} restates the identifying assumptions of \citeasnoun{HuberSolovyeva2017}, under which the direct and indirect effects are point identified, and introduces the sensitivity analysis based on inverse probability weighting when relaxing these assumptions. Section \ref{app} presents an application to the decomposition of the U.S.\ gender wage gap using data from the  National Longitudinal Survey of Youth 1979. Section \ref{concl} concludes.

\section{Variables and parameters of interest}\label{param}

Mediation analysis typically aims to disentangle the average treatment effect (ATE) of a binary treatment, denoted by $D$, on an outcome variable, denoted by $Y$, into a direct effect and an indirect effect operating through one or several mediators. We denote the latter by $M$, which is assumed to have bounded support and may be scalar or a vector of variables and contain discrete and/or continuous elements. For defining natural direct and indirect effects, we make use of the potential outcome framework, see for instance  \citeasnoun{Rubin74}, which has been applied to causal mediation analysis for instance by \citeasnoun{TenHaveetal2007} and \citeasnoun{Albert2008}. Let to this end $M(d), Y(d,M(d'))$ denote the potential mediator state as a function of the treatment and potential outcome as a function of the treatment and the potential mediator, respectively, under treatments $d, d'$ $\in$ $\{0,1\}$. For each subject, only one potential outcome and mediator state, respectively, is observed, because the realized mediator and outcome values are $M=D\cdot M(1) + (1-D)\cdot M(0)$ and $Y=D\cdot Y(1,M(1)) + (1-D)\cdot Y(0,M(0))$.

The ATE, denoted by $\Delta$, is given by the total effect of the treatment operating through the direct or indirect mechanisms:
\begin{eqnarray}
\Delta=E[Y(1,M(1))-Y(0,M(0))].
\end{eqnarray}
The (average) direct effect, denoted by $\theta (d)$, is characterized by the difference in mean potential outcomes under treatment and non-treatment when fixing the mediator at its potential value for $D=d$, which shuts down the indirect mechanism via $M$.
\begin{eqnarray}
\theta (d) &=&E[Y(1,M(d))-Y(0,M(d))],\quad d\in\{0,1\}.
\end{eqnarray}
The (average) indirect effect, denoted by $\delta (d)$,  is given by the difference in mean potential outcomes when exogenously varying the mediator to take its potential values under treatment and non-treatment, but keeping the treatment fixed at $D=d$ to shut down the direct effect.
\begin{eqnarray}
\delta (d) &=&E[Y(d,M(1))-Y(d,M(0))],\quad d\in\{0,1\}.
\end{eqnarray}
\citeasnoun{RoGr92} and \citeasnoun{Robins2003} referred to these causal parameters as pure/total direct and indirect effects, \citeasnoun{FlFl09} as net and mechanism average treatment effects, and \citeasnoun{Pearl01} as natural direct and indirect effects, which is the denomination followed in the remained of this study.

The ATE is the sum of the natural direct and indirect effects defined upon opposite treatment states:
\begin{eqnarray}\label{ate}
\Delta&=&E[Y(1,M(1))-Y(0,M(0))] \notag \\
&=&E[Y(1,M(1))-Y(0,M(1))]+ E[Y(0,M(1))-Y(0,M(0))] = \theta(1) + \delta(0) \notag \\
&=&E[Y(1,M(0))-Y(0,M(0))]+ E[Y(1,M(1))-Y(1,M(0))]=\theta(0) + \delta(1).
\end{eqnarray}
This follows from adding and subtracting either $E[Y(0,M(1))]$ or $E[Y(1,M(0))]$ in \eqref{ate}.The notation $\theta(1),\theta(0)$ and $\delta(1),\delta(0)$ allows for effect heterogeneity as a function of the treatment state, i.e., the presence of interaction effects between the treatment and the mediator. For instance, the impact of a training ($M$) on employment ($Y$) might depend on whether a job seeker has received some form of counseling in the job search process ($D$). A different way to see this is that the direct effect of counseling ($D$) may depend on whether the job seeker attends a training ($M$).

Obviously, effects are not identified without invoking identifying assumptions. First, $Y(1,M(1))$ and $Y(0,M(0))$ are not observed for any subject at the same time, which constitutes the fundamental problem of causal inference. Second, neither $Y(1,M(0))$, nor $Y(0,M(1))$ is observed for any subject. Therefore, point identification of direct and indirect effects requires the treatment and the mediator to be exogenous at least conditional on observables, which appears, however, implausible in many empirical applications. Our sensitivity analysis outlined in Section \ref{sensitivityan} relaxes such exogeneity conditions at the cost of giving up on point identification. This permits incorporating a vector of observed pre-treatment covariates, denoted by $X$, that may confound the causal relations between $D$ and $M$, $D$ and $Y$, and $M$ and $Y$. It is thus assumed that $X$ is insufficient to control for all sources of selection such that unobserved confounders render point identification of direct and indirect effects impossible, which appears plausible in many empirical contexts.

As a further complication to identification, our framework allows for considering outcome attrition/sample selection, implying that $Y$ is only observed for a non-random subpopulation. For instance, when investigating wage outcomes, as in  \citeasnoun{Gr74}, the subpopulation of employed individuals for whom wages are observed might be positively selected in terms of unobservables like ability and motivation. As a further example, consider the effect of educational interventions on test scores, with scores being only observed for those participating in the test or reporting the results, see \citeasnoun{AnBeKr04}. We therefore introduce a binary selection indicator $S$, which indicates whether $Y$ is observed for a specific subject. While $S$ is allowed to be a function of $D$, $M$, and $X$, i.e.\ $S=S(D,M,X)$, it is assumed to neither be affected by nor to affect outcome $Y$. $S$ is therefore not a mediator, as selection per se does not causally influence the outcome, but might nevertheless create endogeneity bias when outcomes are only observed conditional on $S=1$.

\section{Sensitivity analysis}\label{sensitivityan}

The starting point for our sensitivity analysis is a set of assumptions provided in \citeasnoun{HuberSolovyeva2017}, which identifies direct and indirect effects by invoking conditional treatment and mediator exogeneity as well outcome attrition related to observed characteristics (known as missing at random assumption).  Formally, the assumptions are as follows:  \vspace{5pt}\newline
\textbf{Assumption A1 (conditional independence of the treatment):}\newline
(a) $Y(d,m)  \bot D | X=x$, (b) $M(d')  \bot D | X=x$ for all $d,d' \in \{0,1\}$ and $m, x$ in the support of $M, X$.\vspace{5pt}\newline
Assumption A1 rules out unobservables jointly affecting the treatment on the one hand and the mediator and/or the outcome on the other hand conditional on $X$. In contrast, our sensitivity analysis permits that such unobserved confounders do exist. \vspace{5pt}\newline
\textbf{Assumption A2 (conditional independence of the mediator):}\newline
$ Y(d,m) \bot  M | D=d', X=x  $ for all $d,d' \in \{0,1\}$ and $m,x$ in the support of $M,X$. \vspace{5pt}\newline
Assumption A2 rules out unobservables jointly affecting the mediator and the outcome conditional on $D$ and $X$. This only appears plausible if detailed information on possible confounders of the mediator-outcome relation is available in the data (even in experiments with random treatment assignment) and if post-treatment confounders of $M$ and $Y$ can be plausibly ruled out when controlling for $D$ and $X$. In contrast, our sensitivity analysis allows for unobserved confounders of the mediator-outcome relation.
\vspace{5pt}\newline
\textbf{Assumption A3 (conditional independence of selection):}\newline
$ Y \bot  S | D=d, M=m, X=x  $ for all $d \in \{0,1\}$ and $m,x$ in the support of $M,X$. \vspace{5pt}\newline
Assumption A3 rules out unobservables jointly affecting selection and the outcome conditional on $D,M,X$, such that outcomes are missing at random (MAR) in the denomination of \citeasnoun{Ru76b}, i.e.\ outcome attrition is selective w.r.t.\ observed characteristics only. In contrast, our sensitivity analysis permits outcome attrition to be selective w.r.t.\ unobservables.
\vspace{5pt}\newline
\textbf{Assumption A4 (common support):}\newline
(a) $\Pr(D=d| M=m, X=x)>0$ and (b) $\Pr(S=1| D=d, M=m, X=x)>0$ for all $d \in \{0,1\}$ and $m,x$ in the support of $M,X$.\vspace{5pt}\newline
Assumption A4 consists of two common support restrictions. The first requires the conditional probability to receive a specific treatment given $M,X$, henceforth referred to as propensity score, to be larger than zero for either treatment state. This also implies that $\Pr(D=d|X=x)>0$ and (by Bayes' theorem) that $\Pr(M=m| D=d, X=x)>0$, or in the case of $M$ being continuous, that the conditional density of $M$ given $D,X$ is larger than zero. Therefore, $M$ must not be deterministic in $D$ given $X$, as otherwise identification fails due to the lack of comparable units in terms of the mediator across treatment states. The second common support restriction requires that for any combination of $D,M,X$, the probability to be observed is larger than zero. Otherwise, the outcome is not observed for some specific combinations of these variables. Our sensitivity relies on the same set of common support assumptions, in order to make treated and non-treated subjects with observed and non-observed outcomes comparable in terms of observed characteristics.

As outlined in Theorem 1 of \citeasnoun{HuberSolovyeva2017}, Assumptions A1 to A4 permit identifying the mean potential outcomes based on inverse probability weighting (IPW) by
\begin{eqnarray}\label{HS2018T1}
E[ Y(1,M(1))]&=&E\left[ Y\cdot D \cdot S \cdot \frac{1}{\Pr (D=1|X)}\cdot \frac{1}{ \Pr(S=1|D,M,X)} \right], \\
E[ Y(0,M(0))]&=&E\left[ Y\cdot (1-D) \cdot S \cdot \frac{1}{1-\Pr (D=1|X)}\cdot \frac{1}{ \Pr(S=1|D,M,X)} \right], \notag\\
E[ Y(1,M(0))]&=&E\left[ Y\cdot D \cdot S \cdot \frac{1}{1-\Pr(D=1|X)} \cdot \frac{1}{\Pr(S=1|D,M,X)} \cdot \left(\frac{1}{\Pr (D=1|M,X)}-1 \right) \right],\notag\\
E[ Y(0,M(1))]&=&E\left[ Y\cdot (1-D) \cdot S \cdot \frac{1}{\Pr(D=1|X)} \cdot \frac{1}{\Pr(S=1|D,M,X)} \cdot \left(\frac{1}{1-\Pr (D=1|M,X)}-1 \right) \right].\notag
\end{eqnarray}
The direct and indirect effects of interest are obtained as differences between two out of the four mean potential  outcomes. For notational ease, we henceforth denote the various propensity scores in \eqref{HS2018T1} by
\begin{eqnarray}\label{probs}
p^{A1} = \Pr(D=1|X),  \quad  p^{A2} = \Pr(D=1|M,X),  \quad p^{A3}= \Pr(S=1|D,M,X).
\end{eqnarray}
This denomination is motivated by the fact that e.g.\ under A3, there are no confounders that jointly affect $S$ and $D$ conditional on $(M,X)$, $S$ and $M$ conditional on $(D,X)$ or $S$ and $X$ conditional on $(M,D).$ So under A3, $p^{A3}$ is the correct probability to be used for weighting in order to obtain mean potential outcomes. Conversely, if A3 does not hold, e.g. some important confounder is missing, then the correct weight differs from $p^{A3}$.




\tikzstyle{VertexStyle} = [shape            = ellipse,
minimum width    = 6ex,%
draw]

\tikzstyle{EdgeStyle}   = [->,>=stealth']
\begin{figure}[!htp]
	\centering \caption{\label{f1}  Causal paths under conditional exogeneity and missing at random given pre-treatment covariates}\bigskip
	\begin{tikzpicture}[scale=1.5]
	\SetGraphUnit{2}
	\Vertex{D}   \SOEA(D){M} \SO(M){X} \NOEA(M){Y} \SO(Y){S}
	\Edges(X,D,M,Y) \Edges(X,M,S) \Edges(X,Y) \Edges(D,Y)  \Edges(X,S) \Edges(D,S)
	\end{tikzpicture}
\end{figure}
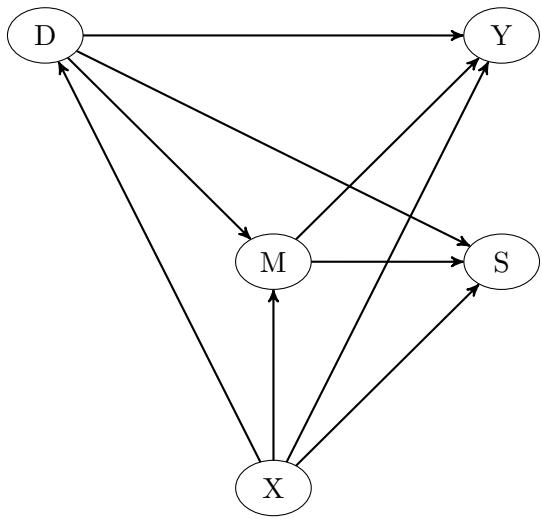


Figure \ref{f1} illustrates our mediation framework with outcome attrition based on a directed acyclic graph, in which the arrows represent causal effects. It is worth noting that each of $D$, $M$, $S$, $X$, and $Y$  might be causally affected by further, unobserved variables not displayed in Figure \ref{f1}. As long as such unobservables do not jointly affect $D$ and $Y$, $D$ and $M$, $M$ and $Y$, or $S$ and $Y$ conditional on $X$, Assumptions A1 to A3 hold. In many empirical problems, however, some or all of A1 to A3 appear difficult defend. While, for instance, A1 holds by design in randomized experiments, it may appear less plausible in observational studies, in particular if the set of observed control variables is limited. Assumption A2 might seem unlikely in any identification design, in particular if there is a substantial time lag between $D$ and $M$ which makes mediator-outcome confounding more likely even when conditioning on pre-treatment covariates $X$. We therefore consider various relaxations of Assumptions A1 to A3.

Our approach modifies the IPW weights of the expressions in \eqref{HS2018T1} to investigate sensitivity and is thus related to  \citeasnoun{HongQinYang2018}, who were the first to propose robustness checks in the context of IPW. However, our approach uses a different measure of discrepancies between the correct and observed weights than \citeasnoun{HongQinYang2018}, whose sensitivity check is based on the covariance between the weight discrepancy and the outcome conditional on the treatment. Also, our approach does not entail analytical formulae and thus relies on optimization routines. Even though this increases the computational burden, an advantage is that we are able to separately consider relaxations of the various identifying assumptions  and thus gain insights on the sources of the potential non-robustness of the effects.

To see the intuition of our approach, suppose for the moment that there exist a scalar of a vector of unobserved confounders, denoted by $U$, that makes Assumption A3 fail. In this case, $p^{A3}$ is no longer the appropriate propensity score for identifying the mean potential outcomes. Our sensitivity analysis is based on
specifying the magnitude by which $p^{A3}$ may deviate from the appropriate (but unidentified) propensity score that includes $U$ as conditioning variable. More formally, we consider the following entropy measure defined as the absolute difference between $p^{A3}$ and the appropriate propensity score for identification, $q^{A3} = \Pr(S=1|D,M,X,U)$:
\begin{eqnarray}\label{relaxation}
|q^{A3} - p^{A3}| &\leq& \epsilon^{A3} \sqrt{p^{A3}(1-p^{A3})}.
\end{eqnarray}
This definition restricts the absolute error in the propensity score $p^{A3}$ due to omitting confounders $U$ to a multiple $ \epsilon^{A3}$ of the standard deviation of a random variable with a binomial distribution corresponding to that of $p^{A3}$. This definition also satisfies a symmetry property, so that the relaxation of the identifying assumption leads to the same set of values for $p^{A3}$ and for $(1-p^{A3}).$ The crucial task is to sensibly choose the value of $\epsilon^{A3}$, e.g.\ based on the richness of $X$, which determines the likely importance of omitted confounders $U$. In an analogous way, $q^{A1} = \Pr(D=1|X,U)$,  $q^{A2} = \Pr(D=1|M,X,U)$ as well as $\epsilon^{A1}$ and $\epsilon^{A2}$ are to be defined.  Suppose there were no unobserved confounders and that Assumptions A1, A2, A3 were satisfied. That would correspond to the situation with $\epsilon^{A1}=\epsilon^{A3}=\epsilon^{A3}=0.$  The greater is the particular entropy parameter, the larger is the permitted importance of unobserved confounders in the specific assumption.

Our sensitivity analysis provides bounds on estimates of the mean potential outcomes in (\ref{HS2018T1}) for deviations of A1, A2, A3 when obeying the restrictions given by $\epsilon^{A1}$, $\epsilon^{A2}$, and $\epsilon^{A3}$ in \eqref{relaxation} as well as specific scaling constraints. The bounds on mean potential outcomes then translate into bounds on natural direct and indirect effects. Assume that we have available an i.i.d.\ sample of $(Y_i,D_i,M_i,X_i,S_i)$ for  $n$ subjects, where $i$ $\in$ $\{1,2,....,n\}$ indexes a specific observation. For the sake of exposition, consider the estimation of $E[Y(1,M(1))]$.  Under Assumptions A1, A3, and A4, this quantity can be estimated by
$$\hat E[ Y(1,M(1))]= \frac{1}{c} \cdot \sum_{i = 1}^{n} Y_i \cdot D_i \cdot S_i \cdot \frac{1}{\hat p_i^{A1}}\cdot \frac{1}{ \hat p_i^{A3}},$$
where $\hat p_i^{A1}, \hat p_i^{A3}$ denote estimates of $p^{A1}, p^{A3}$ for observation $i$, which we obtain in our application presented below by logit regression, and $c$ denotes a normalizing constant, $c = \sum_{i = 1}^{n} \frac{D_i}{\hat p_i^{A1}}\cdot \frac{S_i}{ \hat p_i^{A3}}$.\footnote{Note that Assumption A2 is not required for the identification of $E[Y(d,M(d))]$ for $d \in \{0,1 \}$, but for $E[Y(d,M(1-d))].$}   In the presence of confounders $U$ and a failure of assumptions A1 and/or A3, estimating $\hat E[ Y(1,M(1))]$ based on $\hat p_i^{A1}, \hat p_i^{A3}$ is generally inconsistent. To estimate the bounds on the mean potential outcome under such violations, the unknown population parameters $p^{A1}$ and  $p^{A3}$ are replaced by their estimates $\hat p_i^{A1}$ and $\hat p_i^{A3}$ in  \eqref{relaxation} to estimate the entropy measures $|q^{A1} - p^{A1}|$ and $|q^{A3} - p^{A3}|$, respectively.



Finding the bounds on $\hat E[ Y(1,M(1))]$ corresponds to solving the following optimization problem:
 \begin{eqnarray}
\underset{q^{A1},q^{A3}}{\min/\max} && \frac{1}{c} \cdot \sum_{i = 1}^{n} Y_i \cdot D_i \cdot S_i \cdot \frac{1}{q_i^{A1}}\cdot \frac{1}{ q_i^{A3}} \label{opt}\\
&s.t.& \notag\\
\sum_{i=1}^n \frac{D_i \cdot S_i}{q_i^{A1}} &=&  \sum_{i=1}^n \frac{D_i \cdot S_i}{\hat p_i^{A1}} ,
 \sum_{i=1}^n \frac{D_i \cdot S_i}{q_i^{A3}} = \sum_{i=1}^n \frac{D_i \cdot S_i}{\hat p_i^{A3}} ,
\sum_{i=1}^n \frac{D_i}{q_i^{A1}}\frac{S_i}{q_i^{A3}} =  \sum_{i=1}^n \frac{D_i}{\hat p_i^{A1}} \frac{S_i}{\hat p_i^{A3}} = c   ,    \label{normalize} \\
\forall i: \ |q_i^{A1} - \hat p_i^{A1}| &\leq& \epsilon^{A1,1} \sqrt{\hat p_i^{A1}(1-\hat p_i^{A1})},\quad
|q_i^{A3} - \hat p_i^{A3}| \leq \epsilon^{A3,1} \sqrt{\hat p_i^{A3}(1-\hat p_i^{A3})},\label{relax}\\
q_i^{A1} \in [0,1], && \quad
q_i^{A3} \in [0,1].  \label{probs}
\end{eqnarray}
The equalities in \eqref{normalize} are normalizing restrictions, while the expressions in \eqref{relax} relax the identifying assumptions and \eqref{probs} ensures that $q_i$ are proper probabilities. We note that as only those observations $i$ with $D_i = 1$ and $S_i = 1$ enter the calculations, we added a superscript $1$ to the entropy parameters $\epsilon$ (for $D_i=1$). This implies that one might pick different parameters for different treatment groups, if e.g.\ justified by contextual knowledge.

The optimization problem  \eqref{opt} may be transformed into a computationally more convenient form. Let to this end $\omega_i^{A1} = 1/q_i^{A1}$ and $\omega_i^{A3} = 1/q_i^{A3}$. Then, an alternative representation is
\begin{eqnarray}
\underset{ \omega^{A1},\omega^{A3}}{\min/\max} && \frac{1}{c} \cdot \sum_{i = 1}^{n} Y_i \cdot D_i \cdot S_i \cdot \omega_i^{A1} \cdot \omega_i^{A3}\\
&s.t.& \notag\\
\sum_{i=1}^n D_i \cdot S_i \cdot \omega_i^{A1} &=&  \sum_{i=1}^n \frac{D_i \cdot S_i}{\hat p_i^{A1}} , \quad
 \sum_{i=1}^n  D_i\cdot S_i \cdot \omega_i^{A3} = \sum_{i=1}^n \frac{D_i\cdot S_i } {\hat p_i^{A3}} ,   \notag\\
&&\sum_{i=1}^n  D_i \cdot S_i \cdot \omega_i^{A1} \cdot \omega_i^{A3}=  \sum_{i=1}^n \frac{D_i}{\hat p_i^{A1}} \frac{S_i}{\hat p_i^{A3}} = c   ,   \notag\\
\forall i:  \omega_i^{A1} &\leq& 1/ \left(\hat p_i^{A1} - \epsilon^{A1,1} \sqrt{\hat p_i^{A1}(1-\hat p_i^{A1})} \right),\quad
\omega_i^{A1} \geq 1/ \left(\hat p_i^{A1} + \epsilon^{A1,1} \sqrt{\hat p_i^{A1}(1-\hat p_i^{A1})} \right), \notag\\
\omega_i^{A3} &\leq& 1/ \left(\hat p_i^{A3} - \epsilon^{A3,1} \sqrt{\hat p_i^{A3}(1-\hat p_i^{A3})} \right), \quad
\omega_i^{A3} \geq 1/ \left(\hat p_i^{A3} + \epsilon^{A3,1} \sqrt{\hat p_i^{A3}(1-\hat p_i^{A3})} \right), \notag\\
\omega_i^{A1} &\geq& 0,  \quad
\omega_i^{A3} \geq 0.   \notag
\end{eqnarray}
For a fixed vector $\omega^{A1}$ or $\omega^{A3}$, this problem is a linear program. This suggests an algorithm that iteratively changes $\omega^{A1}$ and $\omega^{A2}$ until convergence using $\omega_i^{A1} = 1/\hat p_i^{A1}$ and $\omega_i^{A3} = 1/\hat p_i^{A3}$ as starting point.
An important feature of our approach based on the optimization is that we impose no structure on the dependence of potentially omitted confounders and our outcome variable and exhaust all possibilities for the weights. This may in general lead to wider bounds and thus more prudent inference.

Bounds on  $\hat E[Y(0,M(0))]$ (i.e.\ the estimate of $E[Y(0,M(0))]$) can be constructed in an analogous manner by using observations with $S_i=1$, $D_i=0$ and searching through $(1-q_i^{A1})$ instead. For bounds on $\hat E[Y(1,M(0))]$ and $\hat E[Y(0,M(1))]$, one has to optimize w.r.t.\ $q_i^{A1}$, $q_i^{A2}$ and $q_i^{A3}.$ All optimization problems are formally stated in Appendix \ref{bounds}.
After deriving upper and lower bounds on the mean potential outcomes,  we may construct bounds on the effects of interest in the following way:
\begin{eqnarray}\label{MTO2}
\hat \Delta_{LB} &=&\hat E[ Y(1,M(1))]_{LB} - \hat E[ Y(0,M(0))]_{UB}, \notag\\
\hat \Delta_{UB} &=& \hat E[ Y(1,M(1))]_{UB} - \hat E[ Y(0,M(0))]_{LB}, \notag\\
\hat \theta(1)_{LB} &=&\hat E[ Y(1,M(1))]_{LB} - \hat E[ Y(0,M(1))]_{UB}, \notag\\
\hat \theta(1)_{UB} &=& \hat E[ Y(1,M(1))]_{UB} - \hat E[ Y(0,M(1))]_{LB}, \notag\\
\hat \theta(0)_{LB} &=&\hat E[ Y(1,M(0))]_{LB} - \hat E[ Y(0,M(0))]_{UB}, \notag\\
\hat \theta(0)_{UB} &=&\hat E[ Y(1,M(0))]_{UB} - \hat E[ Y(0,M(0))]_{LB}, \notag\\
\hat \delta(1)_{LB} &=&\hat E[ Y(1,M(1))]_{LB} - \hat E[ Y(1,M(0))]_{UB}, \notag\\
\hat \delta(1)_{UB} &=&\hat E[ Y(1,M(1))]_{UB} - \hat E[ Y(1,M(0))]_{LB}, \notag\\
\hat \delta(0)_{LB} &=&\hat E[ Y(0,M(1))]_{LB} - \hat E[ Y(0,M(0))]_{UB}, \notag\\
\hat \delta(0)_{UB} &=&\hat E[ Y(0,M(1))]_{UB} - \hat E[ Y(0,M(0))]_{LB}, \notag
\end{eqnarray}
where subscripts $LB, UB$  stand for the lower  and upper bounds of the respective estimated mean potential outcome and `$\hat{\ }$' implies that any of the causal effects are estimates rather than population parameters. The bounds on $\Delta$, $\theta(1)$, and $\theta(0)$ are sharp,
because  the observations that are used for calculating the two mean potential outcomes upon which the respective effect is defined are distinct. Consider for instance the lower bound on $\theta(1)$. In order to estimate $E[ Y(1,M(1))]_{LB}$, we use observations with $D_i=1$, while for $E[ Y(0,M(1))]_{UB}$ we only use observations with $D_i=0$. In contrast, the bounds for $\delta(1)$ and $\delta(0)$ are not necessarily sharp. As an example, consider $\delta(1)$. In order to estimate $E[ Y(1,M(1))]_{LB}$ and $E[ Y(1,M(0))]_{UB}$, we use observations with $D_i=1$ and the weights that entail $E[ Y(1,M(1))]_{LB}$ and $E[ Y(1,M(0))]_{UB}$ are not necessary the same.\footnote{ It is in principle possible to compute sharp bounds even for $\delta(1)$ and $\delta(0)$ at the cost of solving a more complex optimization problem. In such case, however, our heuristic algorithm of solving sequential linear programs cannot be used.}

An important question yet to be discussed is how to set the entropy parameters, which represent changes in the propensity scores due to omitting confounders (e.g.\ $\epsilon^{A1,1}$ and $\epsilon^{A3,1}$), in a meaningful way. Investigating the importance of observed confounders $X$ may provide some guidance for finding sensible values for the entropy parameters. Consider, for instance, a logistic regression for estimating $p_i^{A3}$:
\begin{eqnarray*}\label{log}
S_i &\sim& Bern(p_i^{A3}), \\
\log\left( \frac{p_i^{A3}}{1-p_i^{A3}} \right) &=& \alpha_0 + \alpha_D D_i + \alpha_M^T M_i + \alpha_X^T X_i,
\end{eqnarray*}
where  $\hat p_i^{A3}$ is obtained by maximum likelihood estimation. Now consider removing the most important predictor (of $S$) in $X$ and re-estimating the propensity score, denoted as $\hat p_{i,X1}^{A3}$, where $X1$ are the remaining covariates (without the most important predictor).

For each observation, one can then compute the entropy parameter when including and excluding the most important predictor in $X$,  $\epsilon_{i,X1}^{A3} = \frac{|\hat p_{i,X1}^{A3} - \hat p_i^{A3}|}{ \sqrt{\hat p_i^{A3}(1-\hat p_i^{A3})}}$. One may ultimately pick the entropy parameter as average of $\epsilon_{i,X1}^{A3}$ in the subpopulation with $D_i=1$ and $S_i=1$:
\begin{eqnarray*}\epsilon^{A3,1}_{X1} = \sum_{i=1}^n \frac{D_i \cdot S_i  \cdot \epsilon_{i,X1}^{A3}}{\sum_{i=1}^n D_i \cdot S_i }.\end{eqnarray*}
This corresponds to the average change induced by omitting the most important predictor from $X$, which is used as a proxy for the importance of unobserved confounders $U$. There are different ways of measuring the importance of a predictor in a regression and natural choice seems to be the change in the deviance. The latter is the log-likelihood ratio statistic for testing the difference in the model fit when including and excluding a specific predictor, which is asymptotically chi-squared distributed. Similarly $\epsilon^{A3}_{Xj}$ and $\epsilon^{A3}_{Mj}$ would denote the average change in estimated probabilities that would omitting the $j$-th most important (measured the by  deviance) from $X$ and $M$, respectively.\footnote{Another approach for setting the entropy parameter is to consider the change in estimated probabilities induced by a change of the link function, e.g.\ by picking the probit instead of logit function. Formally, $\epsilon^{A3,1}_{i,probit}  = \frac{|\hat p_{i,probit}^{A3} - \hat p_i^{A3}|}{ \sqrt{\hat p_i^{A3}(1-\hat p_i^{A3})}}$, where $\hat p_{i,probit}^{A3}$ and $\hat p_i^{A3}$ correspond to the estimated probabilites under a probit and logit model, respectively. One may thus pick the entropy parameter as average
$\epsilon^{A3,1}_{probit} = \sum_{i=1}^n \frac{D_i \cdot S_i  \cdot \epsilon^{A3}_{i,probit} }{\sum_{i=1}^n D_i \cdot S_i }.$}

\section{Application}\label{app}

We apply our method to data from the National Longitudinal Survey of Youth 1979 (NLSY79), a panel survey of young individuals who were aged 14 to 22 years in 1979, to decompose the gender wage gap in the year 2000 when respondents were 35 to 43 years old.\footnote{The NLSY79 data consists of three  independent samples: a cross-sectional sample (6,111 subjects, or 48\%)  representing the non-institutionalized civilian youth; a supplemental sample (42\%) oversampling civilian Hispanic, black, and economically disadvantaged nonblack/non-Hispanic young people; and a military sample (10\%) comprised of youth serving in the military as of September 30, 1978 (\citeasnoun{NLSY792000}).} We use exactly the same sample definition as in \citeasnoun{HuberSolovyeva2019}, who consider five wage decomposition techniques with distinct identifying assumptions to investigate the sensitivity of point estimators of the indirect effect (or explained component) due to gender differences in mediators like education or occupation as well as the direct gender effect (or unexplained component). However, the consistency of these decomposition techniques relies on specific conditional exogeneity or instrumental variable assumptions w.r.t.\ to gender, the mediators, and the observability of wages, which are likely violated in practice. See also \citeasnoun{Huber2015} for a discussion of identification issues with in wage decompositions based on the causal mediation framework.  In contrast, the approach suggested in this paper permits investigating the robustness of the direct and indirect effects of gender on wage under violations of conditional exogeneity.

The NLSY79 includes a rich set of individual characteristics, including socio-economic variables likes education, occupation, work experience, and a range of further labor market-relevant information. Our evaluation sample consists of 6,658 individuals (3,162 men and 3,496 women), after excluding 1,351 observations from the total NLSY79 sample in 2000 due to various data issues outlined in \citeasnoun{HuberSolovyeva2019}.\footnote{For instance, we excluded 502 persons reporting to work 1,000 hours or more in the past calendar year, but whose average hourly wages in the past calendar year were either missing or equal to zero. Furthermore, we dropped 54 working individuals with average hourly wages of less than \$1 in the past calendar year. We also excluded 608 observations with missing values in the mediators.}  Treatment $D$ is a binary indicator for gender, taking the value zero for females and one for males. Outcome $Y$ corresponds to the log average hourly wage in the past calendar year reported in 2000. However, the wage outcome under full-time employment is not observed
for everyone, as a non-negligible share (in particular among females) are in minor employment or not employed at all. This likely introduces sample selection issues, see \citeasnoun{Heckman79}. We therefore define the selection indicator $S$ to be one for individuals who indicated to have worked at least 1,000 hours in the past calendar year, as it is the case for 87\% of males and 70\% of females.

The vector of mediators $M$ includes individual variables reported in or constructed with reference to 1998 such that they arguably reflect decisions taken after birth and prior to the measurement of the outcome (i.e.\ on the causal path between $D$ and $Y$): marital status, years in marriage, the region and the number of years residing in that region, a dummy for living in an urban area (SMSA) and the number of years living in an urban area, education, dummies for the year of first employment, number of jobs ever had, tenure with the current employer (in weeks), industry and the number of years working in that industry, occupation and the number of years working in that occupation, whether employed in 1998 and total years of employment, a dummy for full-time employment and the share of full-time employment from 1994-98, total weeks of employment, the number of weeks unemployed and the number of weeks out of the labor force, and whether health problems prevented employment along with the history of health problems.

In the propensity scores, we control for a set of observed covariates $X$ arguably mostly determined at or prior to birth, namely race, religion, year of birth, birth order, parental place of birth (in the U.S. or abroad), and parental education. However, unobserved confounders causing non-ignorable selection into the treatment, mediators, and/or employment decision $S$  even after controlling for $X$ likely exist. For instance, risk preferences, attitudes towards competition and negotiations, and other socio-psychological factors, see e.g., \citeasnoun{Bertrand} and \citeasnoun{AzmatPet14}, are not available in our data. Such variables might possibly confound the mediator-outcome relationship. As a second example, selection into employment might be driven by innate ability or motivation, which likely also affect wages.  Due to such endogeneity concerns, one should be cautious about deriving causal claims (e.g. about the amount of labor market discrimination) and policy recommendations based on wage gap decompositions, see the discussion in \citeasnoun{HuberSolovyeva2019}. In this context, our method is useful for assessing the sensitivity of the results to violations of some or all exogeneity conditions required for a causal interpretation of wage decompositions.

Table \ref{tab:descr} in Appendix \ref{app:app} provides descriptive statistics for the key variables of our analysis, namely means, mean differences across gender, and the respective \textit{p}-values based on two-sample \textit{t}-tests. The \textit{p}-values suggest that females an males differ importantly in a range of variables like labor market experience,  education, industry, and occupation. Our sensitivity analysis is based on assuming that omitted confounders in the various propensity score specifications behave similar to the first, second, or third important predictors among the pre-treatment covariates or mediators that enter the respective specification. For this reason, Table  \ref{tab:imp}  shows the three most important  covariates and mediators in the different propensity score models presumably prone to confounding, where importance refers to the change in deviance as discussed at the end of Section \ref{sensitivityan}.

Tables \ \ref{tab1}, \ref{tab2}, \ref{tab3}, and \ref{tab4} \ present the estimated effect bounds on $\theta(1)$, $\theta(0)$,  $\delta(1)$, and $\delta(0)$, respectively, in squared brackets, when basing the entropy parameter on the respective first, second, or third most important predictors.\footnote{We note that in the decomposition literature, it is frequently the male wages that are considered as reference (or `fair') wages. This suggests considering $\theta(0)$ and   $\delta(1)$ as unexplained and explained components, respectively. See \citeasnoun{Sloczynski2013} for an in-depth discussion of reference group choice in the potential outcome framework.} 95\% confidence intervals are also reported in parentheses, based on subsampling with 500 replications and a subsample size of $\floor{n^{0.7}}$.\footnote{We applied subsampling to the upper and lower bounds separately similar to \citeasnoun{laffers2017sensitivity} or \citeasnoun{demuynck2015bounding}. A computationally more expensive stepdown procedure of \citeasnoun{romano2010inference} may be used to control for the asymptotic coverage of the whole identified set. }We observe that the direct effects remain statistically significant at the 5\% level under most relaxations and that potentially missing confounders would have the biggest impact via Assumption A2. Previous employment has considerable explanatory power for later labor market performance, as reflected in the relatively wide bounds in the column for the 2nd most important missing $M$ (employment in 1998) in Table \ref{tab1}, where the lower 95\% confidence bound goes below zero. Table \ref{tab2} displays the estimated bounds for the natural direct effect for $d=0$, when mediators are set to their potential values for women and the indirect mechanisms are shut down. For this effect, the choice of the link function seems important when considering Assumption A2 and the related regression for estimating $P(D=1|M,X).$ More precisely, if we would allow the difference between the correct and estimated weights to be of a similar magnitude as is the difference from using the probit instead of logit link function, then the upper bound on this effect may exceed $0.6$.

Concerning the natural indirect effects reported in Tables \ref{tab3} and \ref{tab4}, the confidence interval on the effect estimate of $0.053$ includes the zero under most relaxations for males ($d=1$), while the lower confidence bound remains above zero under most relaxations for females ($d=0$). We also observe that confidence intervals are not necessarily symmetric around the estimated bounds and that the omission of the variable with the most predictive power (measured by the change in deviance) does not necessarily lead to the widest bounds. The latter is due to the fact that e.g.\ the strongest predictor of the treatment is not necessarily the strongest confounder, i.e.\ the predictor's association with the outcome is sufficiently weaker than that of other treatment predictors.

\bigskip

\begin{table}[h!]
{\footnotesize
\begin{center}
\caption{Covariates and mediators with the highest predictive power} \label{tab:imp}
\vspace{2ex}
\begin{tabular}{ ccl } 
\hline
\hline
\it{ Assumption A1 } & \multicolumn{2}{c}{$P(D=1|X)$} \vspace{1ex} \\
Most important $X$ 	& 1st & Mothers educ. missing \\
 		 		& 2nd &  Mothers educ. high school graduate  \\
	   		  	& 3rd & Religion missing \vspace{1ex} \\
\hline
 \it{ Assumption A2 } & \multicolumn{2}{c}{$P(D=1|M,X)$}\vspace{1ex} \\
Most important $M$ 	& 1st & Farmer or laborer \\
 		 		& 2nd & Industry: Professional services  \\
	   		  	& 3rd & Clerical occupation \vspace{1ex} \\
Most important $X$ 	& 1st & White \\
 		 		& 2nd & Fathers educ. college/more  \\
	   		  	& 3rd & Mothers educ. missing \vspace{1ex} \\
\hline
 \it{ Assumption A3 } & \multicolumn{2}{c}{$P(S=1|D,M,X)$}\vspace{1ex} \\
Most important $M$ 	& 1st & Employed full time \\
 		 		& 2nd & Employment status: employed  \\
	   		  	& 3rd & Operator (machines, transport) \vspace{1ex} \\
Most important $X$ 	& 1st & Fathers educ. college/more \\
 		 		& 2nd & Mothers educ. some college  \\
	   		  	& 3rd & Protestant  \vspace{1ex} \\
\hline
\end{tabular}
\end{center}
\par
Note: Most important predictors in different propensity score estimations measured by the change in deviance.
}
\end{table}

\clearpage

\begin{table}[h!]
{\footnotesize
\begin{center}
\caption{Bounds on the natural direct effect under $d=1$ (point estimate: 0.201)} \label{tab1}
\vspace{2ex}
\begin{tabular}{ c  ccc  c} 
\hline
\hline
  \multirow{2}{*}{   Importance }   & \multicolumn{3}{c}{  Missing $X$}   & \multirow{2}{*}{   Probit   }\\
&    1st  &   2nd   &   3rd   &     \\
\hline
\multirow{2}{*}{A1} & [ 0.192   0.210 ] & [ 0.184   0.217 ] & [ 0.197   0.205 ] & [ 0.201   0.201 )  \\
& ( 0.105   0.303 ) & ( 0.098   0.310 ) & ( 0.110   0.298 ) & ( 0.114   0.294 )  \\
\multirow{2}{*}{A2} & [ 0.165   0.237 ] & [ 0.166   0.235 ] & [ 0.181   0.220 ]  & [ 0.177   0.291 ]  \\
& ( 0.080   0.327 ) & ( 0.081   0.325 ) & ( 0.095   0.311 )  & ( 0.091   0.381 )  \\
\multirow{2}{*}{A3} & [ 0.190   0.211 ] & [ 0.192   0.209 ] & [ 0.192   0.209 & [ 0.191   0.241 ]  \\
& ( 0.104   0.305 ) & ( 0.106   0.303 ) & ( 0.106   0.302 ) & ( 0.104   0.339 )  \\
\multirow{2}{*}{A1 + A2} & [ 0.156   0.246 ] & [ 0.150   0.252 ] & [ 0.177   0.224 ] & [ 0.114   0.291 ]  \\
& ( 0.071   0.336 ) & ( 0.065   0.342 ) & ( 0.091   0.315 ) &  ( 0.034   0.381 )  \\
\multirow{2}{*}{A2 + A3} & [ 0.154   0.248 ] & [ 0.158   0.244 ] & [ 0.173   0.229 ]  & [ 0.167   0.332 ]  \\
& ( 0.069   0.338 ) & ( 0.073   0.335 ) & ( 0.087   0.320 ) & ( 0.081   0.426 )  \\
\multirow{2}{*}{A1 + A3} & [ 0.196   0.205 ] & [ 0.197   0.205 ] & [ 0.197   0.204 ] & [ 0.188   0.213 ]  \\
& ( 0.111   0.298 ) & ( 0.111   0.298 ) & ( 0.111   0.298 )  & ( 0.112   0.305 )  \\
\multirow{2}{*}{A1 + A2 + A3} & [ 0.145   0.257 ] & [ 0.142   0.260 ] & [ 0.169   0.233 ]  & [ 0.075   0.332 ]  \\
& ( 0.060   0.348 ) & ( 0.057   0.352 ) & ( 0.083   0.324 ) & ( -0.006   0.426 )  \\
\hline
\noalign{\smallskip}	
  \multirow{2}{*}{   Importance }   &  \multicolumn{3}{c}{  Missing $M$}  & \multirow{2}{*}{      }\\
  &    1st    &   2nd   &   3rd  &  \\
\hline
\multirow{2}{*}{A2} &  [ 0.114   0.237 ] & [ 0.068   0.235 ] & [ 0.082   0.220 ] &  \\
 & ( 0.034   0.327 ) & ( -0.016   0.325 ) & ( 0.003   0.311 ) & \\
\multirow{2}{*}{A3} & [ 0.162   0.211 ] & [ 0.172   0.209 ] & [ 0.181   0.209 ] &  \\
& ( 0.080   0.305 ) & ( 0.086   0.303 ) & ( 0.095   0.302 ) &   \\
\multirow{2}{*}{A2 + A3} & [ 0.075   0.248 ] & [ 0.039   0.244 ] & [ 0.063   0.229 ] &  \\
& ( -0.006   0.338 ) & ( -0.045   0.335 ) & ( -0.019   0.320 ) &  \\
 \hline
\end{tabular}
\end{center}
\par
Note: Estimated bounds on $\theta(1)$  using entropy parameters corresponding to the average change induced by omitting the 1st, 2nd and the 3rd most important predictor in $X$ or $M$ (listed in Table \ref{tab:imp}). 95\% confidence intervals based on subsampling with 500 replications and subsample size $\floor{n^{0.7}}$ are in parenthesis. The last column corresponds to the change due to a different choice of the link function, namely probit instead of logit.
}
\end{table}

\clearpage

\begin{table}[h!]
{\footnotesize
\begin{center}
\caption{Bounds on the natural direct effect under $d=0$ (point estimate: 0.325)} \label{tab2}
\vspace{2ex}
\begin{tabular}{ c  ccc c} 
\hline
\hline
  \multirow{2}{*}{   Importance }   & \multicolumn{3}{c}{  Missing $X$}    & \multirow{2}{*}{   Probit   }\\
 &    1st  &   2nd   &   3rd &  \\
\hline
\multirow{2}{*}{A1} & [ 0.308   0.342 ] & [ 0.295   0.356 ] & [ 0.319   0.332 ] & [ 0.325   0.326 )  \\
& ( 0.202   0.457 ) & ( 0.188   0.469 ) & ( 0.213   0.447 ) &  ( 0.219   0.441 )  \\
\multirow{2}{*}{A2} & [ 0.282   0.366 ] & [ 0.285   0.364 ] & [ 0.303   0.347 ] ] & [ 0.299   0.475 ]  \\
& ( 0.174   0.479 ) & ( 0.177   0.477 ) & ( 0.196   0.461 )  & ( 0.192   0.586 )  \\
\multirow{2}{*}{A3} & [ 0.308   0.344 ] & [ 0.311   0.340 ] & [ 0.312   0.340 ] & [ 0.309   0.405 ]  \\
& ( 0.201   0.461 ) & ( 0.204   0.457 ) & ( 0.205   0.456 ) & ( 0.202   0.570 )  \\
\multirow{2}{*}{A1 + A2} & [ 0.266   0.383 ] & [ 0.255   0.394 ] & [ 0.297   0.353 ] & [ 0.121   0.475 ]  \\
& ( 0.157   0.495 ) & ( 0.146   0.505 ) & ( 0.189   0.467 ) & ( -0.011   0.586 )  \\
\multirow{2}{*}{A2 + A3} & [ 0.265   0.384 ] & [ 0.270   0.379 ] & [ 0.289   0.361 ] & [ 0.283   0.554 ]  \\
& ( 0.155   0.499 ) & ( 0.161   0.493 ) & ( 0.181   0.476 ) & ( 0.175   0.711 )  \\
\multirow{2}{*}{A1 + A3} & [ 0.311   0.339 ] & [ 0.305   0.345 ] & [ 0.317   0.333 ] &  [ 0.311   0.338 ]  \\
& ( 0.205   0.453 ) & ( 0.198   0.457 ) & ( 0.211   0.448 ) & ( 0.204   0.453 )  \\
\multirow{2}{*}{A1 + A2 + A3} & [ 0.248   0.401 ] & [ 0.240   0.409 ] & [ 0.283   0.368 ] & [ 0.055   0.554 ]  \\
& ( 0.138   0.514 ) & ( 0.130   0.521 ) & ( 0.175   0.482 ) & ( -0.080   0.698 )  \\
\hline
\noalign{\smallskip}	
  \multirow{2}{*}{   Importance }   & \multicolumn{3}{c}{  Missing $M$}  & \multirow{2}{*}{     }\\
 &    1st  &   2nd   &   3rd     \\
\hline
\multirow{2}{*}{A2} & [ 0.121   0.366 ] & [ 0.142   0.364 ] & [ 0.203   0.347 ] & \\
& ( -0.011   0.479 ) & ( 0.014   0.477 ) & ( 0.085   0.461 ) &  \\
\multirow{2}{*}{A3} & [ 0.261   0.344 ] & [ 0.278   0.340 ] & [ 0.293   0.340 ] & \\
& ( 0.152   0.461 ) & ( 0.167   0.457 ) & ( 0.185   0.456 ) &  \\
\multirow{2}{*}{A2 + A3} & [ 0.055   0.384 ] & [ 0.094   0.379 ] & [ 0.171   0.361 ] &   \\
 & ( -0.079   0.499 ) & ( -0.036   0.493 ) & ( 0.050   0.476 ) &   \\
\hline
\end{tabular}
\end{center}
\par
Note: Estimated bounds on $\theta(0)$ using entropy parameters corresponding to the average change induced by omitting the 1st, 2nd and the 3rd most important predictor in $X$ or $M$ (listed in Table \ref{tab:imp}). 95\% confidence intervals based on subsampling with 500 replications and subsample size $\floor{n^{0.7}}$ are in parenthesis. The last column corresponds to the change due to a different choice of the link function, namely probit instead of logit.
}
\end{table}

\clearpage

\begin{table}[h!]
{\footnotesize
\begin{center}
\caption{Bounds on the natural indirect effect under $d=1$ (point estimate: 0.053)} \label{tab3}
\vspace{2ex}
\begin{tabular}{ c  ccc c} 
\hline
\hline
  \multirow{2}{*}{   Importance }   & \multicolumn{3}{c}{  Missing $X$}   & \multirow{2}{*}{   Probit   }\\
&    1st  &   2nd   &   3rd   &  \\
\hline
\multirow{2}{*}{A1} & [ 0.035   0.071 ] & [ 0.021   0.085 ] & [ 0.045   0.061 ] & [ 0.053   0.053 )  \\
& ( -0.067   0.145 ) & ( -0.081   0.159 ) & ( -0.057   0.135 ) &  ( -0.049   0.127 )  \\
\multirow{2}{*}{A2} & [ 0.012   0.096 ] & [ 0.015   0.093 ] & [ 0.031   0.075 ] & [ 0.028   0.257 ]  \\
& ( -0.084   0.172 ) & ( -0.082   0.170 ) & ( -0.068   0.150 ) & ( -0.071   0.353 )  \\
\multirow{2}{*}{A3} & [ 0.044   0.062 ] & [ 0.045   0.060 ] & [ 0.046   0.060 ] & [ 0.043   0.076 ]  \\
& ( -0.059   0.137 ) & ( -0.058   0.136 ) & ( -0.057   0.135 ) & ( -0.060   0.154 )  \\
\multirow{2}{*}{A1 + A2} & [ -0.006   0.113 ] & [ -0.018   0.124 ] & [ 0.023   0.083 ] & [ -0.097   0.257 ]  \\
& ( -0.102   0.190 ) & ( -0.113   0.201 ) & ( -0.076   0.158 ) & ( -0.183   0.353 )  \\
\multirow{2}{*}{A2 + A3} & [ 0.003   0.105 ] & [ 0.007   0.101 ] & [ 0.024   0.082 ] & [ 0.018   0.282 ]  \\
& ( -0.093   0.183 ) & ( -0.089   0.178 ) & ( -0.075   0.159 ) &  ( -0.081   0.380 )  \\
\multirow{2}{*}{A1 + A3} & [ 0.039   0.067 ] & [ 0.033   0.073 ] & [ 0.045   0.061 ] & [ 0.040   0.067 ]  \\
& ( -0.063   0.142 ) & ( -0.069   0.148 ) & ( -0.057   0.136 ) &  ( -0.064   0.143 )  \\
\multirow{2}{*}{A1 + A2 + A3} & [ -0.015   0.123 ] & [ -0.025   0.132 ] & [ 0.016   0.090 ] & [ -0.120   0.282 ]  \\
& ( -0.110   0.200 ) & ( -0.120   0.210 ) & ( -0.083   0.167 ) &  ( -0.205   0.380 )  \\
\hline
\noalign{\smallskip}	
  \multirow{2}{*}{   Importance }   &  \multicolumn{3}{c}{  Missing $M$}  & \multirow{2}{*}{   }\\
&    1st  &   2nd   &   3rd   &    \\
\hline
\multirow{2}{*}{A2} &[ -0.097   0.096 ] & [ -0.087   0.093 ] & [ -0.051   0.075 ] & \\
 & ( -0.183   0.172 ) & ( -0.174   0.170 ) & ( -0.142   0.150 ) & \\
\multirow{2}{*}{A3} & [ 0.029   0.062 ] & [ 0.032   0.060 ] & [ 0.032   0.060 ] &  \\
 & ( -0.076   0.137 ) & ( -0.073   0.136 ) & ( -0.072   0.135 ) &  \\
\multirow{2}{*}{A2 + A3}& [ -0.120   0.105 ] & [ -0.108   0.101 ] & [ -0.071   0.082 ] &  \\
 & ( -0.205   0.183 ) & ( -0.194   0.178 ) & ( -0.162   0.159 ) & \\
\hline

\end{tabular}
\end{center}
\par
Note: Estimated bounds on $\delta(1)$  using entropy parameters corresponding to the average change induced by omitting the 1st, 2nd and the 3rd most important predictor in $X$ or $M$ (listed in Table \ref{tab:imp}). 95\% confidence intervals based on subsampling with 500 replications and subsample size $\floor{n^{0.7}}$ are in parenthesis. The last column corresponds to the change due to a different choice of the link function, namely probit instead of logit.
}
\end{table}

\clearpage

\begin{table}[h!]
{\footnotesize
\begin{center}
\caption{Bounds on the natural indirect effect under $d=0$ (point estimate: 0.177)} \label{tab4}
\vspace{2ex}
\begin{tabular}{ c ccc c} 
\hline
\hline
  \multirow{2}{*}{  Importance  } & \multicolumn{3}{c}{ Missing X } & \multirow{2}{*}{ Probit  }\\
&  1st  & 2nd  & 3rd  &  \\
\hline
\multirow{2}{*}{A1} & [ 0.169   0.186 ] & [ 0.163   0.192 ] & [ 0.175   0.180 ] &  [ 0.177   0.178 )  \\
& ( 0.063   0.272 ) & ( 0.056   0.279 ) & ( 0.068   0.267 ) &  ( 0.071   0.264 )  \\
\multirow{2}{*}{A2} & [ 0.141   0.213 ] & [ 0.143   0.212 ] & [ 0.158   0.197 ] &  [ 0.153   0.264 ]  \\
& ( 0.039   0.299 ) & ( 0.041   0.297 ) & ( 0.055   0.283 ) &  ( 0.051   0.348 )  \\
\multirow{2}{*}{A3} & [ 0.158   0.197 ] & [ 0.162   0.194 ] & [ 0.163   0.193 ] &  [ 0.161   0.272 ]  \\
& ( 0.050   0.286 ) & ( 0.054   0.282 ) & ( 0.055   0.281 ) & ( 0.053   0.396 )  \\
\multirow{2}{*}{A1 + A2} & [ 0.133   0.222 ] & [ 0.128   0.227 ] & [ 0.155   0.200 ] & [ 0.087   0.264 ]  \\
& ( 0.031   0.307 ) & ( 0.026   0.313 ) & ( 0.052   0.286 ) &( -0.011   0.348 )  \\
\multirow{2}{*}{A2 + A3} & [ 0.122   0.233 ] & [ 0.127   0.228 ] & [ 0.143   0.213 ] &  [ 0.137   0.359 ]  \\
& ( 0.020   0.322 ) & ( 0.025   0.316 ) & ( 0.039   0.301 ) & ( 0.034   0.480 )  \\
\multirow{2}{*}{A1 + A3} & [ 0.173   0.182 ] & [ 0.173   0.181 ] & [ 0.174   0.181 ] & [ 0.165   0.190 ]  \\
& ( 0.066   0.269 ) & ( 0.067   0.268 ) & ( 0.067   0.268 ) &  ( 0.054   0.275 )  \\
\multirow{2}{*}{A1 + A2 + A3} & [ 0.114   0.241 ] & [ 0.112   0.242 ] & [ 0.140   0.215 ] & [ 0.005   0.359 ]  \\
& ( 0.012   0.330 ) & ( 0.010   0.331 ) & ( 0.036   0.303 ) & ( -0.096   0.477 )  \\
\hline
\noalign{\smallskip}	
  \multirow{2}{*}{Importance  } &  \multicolumn{3}{c}{ Missing M} & \multirow{2}{*}{ \textbf{ } }\\
&  1st  & 2nd  & 3rd  &  \\
\hline
\multirow{2}{*}{A2}& [ 0.087   0.213 ] & [ 0.035   0.212 ] & [ 0.052   0.197 ] & \\
& ( -0.011   0.299 ) & ( -0.071   0.297 ) & ( -0.044   0.283 ) &   \\
\multirow{2}{*}{A3} & [ 0.096   0.197 ] & [ 0.122   0.194 ] & [ 0.146   0.193 ] & \\
& ( -0.008   0.286 ) & ( 0.011   0.282 ) & ( 0.037   0.281 ) &  \\
\multirow{2}{*}{A2 + A3}  & [ 0.005   0.233 ] & [ -0.021   0.228 ] & [ 0.021   0.213 ] & \\
& ( -0.090   0.322 ) & ( -0.125   0.316 ) & ( -0.077   0.301 ) &  \\

\hline

\end{tabular}
\end{center}
\par
Note: Estimated bounds on $\delta(0)$  using entropy parameters corresponding to the average change induced by omitting the 1st, 2nd and the 3rd most important predictor in $X$ or $M$ (listed in Table \ref{tab:imp}). 95\% confidence intervals based on subsampling with 500 replications and subsample size $\floor{n^{0.7}}$ are in parenthesis. The last column corresponds to the change due to a different choice of the link function, namely probit instead of logit.
}
\end{table}

\section{Conclusion}\label{concl}

This paper proposed a sensitivity check for estimating natural direct and indirect effects in the presence of treatment and mediator endogeneity as well as selective attrition or missingness in outcomes. To this end, we considered identification based on inverse probability weighting using treatment and selection propensity scores and perturbed the respective propensity scores by an entropy parameter reflecting a specific amount of misspecification to set-identify the effects of interest. We demonstrated that this approach can be framed as a linear programming problem and discussed sensible choices of the entropy parameters based on the predictive power of the most important predictors in the propensity scores. Finally, we applied our method to data from the National Longitudinal Survey of Youth 1979 to derive bounds on the explained and unexplained components of a gender wage gap decomposition that is likely prone to non-ignorable mediator selection and sample selection in terms of the observability of the wage outcomes.

{\footnotesize
	\bibliographystyle{econometrica}
	\bibliography{research}}

@ARTICLE{Sloczynski2013,
  author =       {T Sloczynski},
  title =        {Population average gender effects},
  journal =      {IZA Discussion Paper No. 7315},
  year =         {2013},
}

@ARTICLE{HuberSolovyeva2019,
	author =       {Martin Huber and Anna Solovyeva},
	title =        {On the sensitivity of wage gap decompositions},
	journal =      {forthcoming in the Journal of Labor Research},
	year =         {2020},
}

@ARTICLE{HongQinYang2018,
  author =       {Guanglei Hong and Xu Qin and Fan Yang,},
  title =        {Weighting-Based Sensitivity Analysis in Causal Mediation Studies},
  journal =      {Journal of Educational and Behavioral Statistics},
  year =         {2018},
  volume =       {43},
  pages =        {32-56},
}

@ARTICLE{VansteelandtVanderWeele2012,
  author =       {Vansteelandt, S. and VanderWeele, T. J.},
  title =        {Natural direct and indirect effects on the exposed: effect decomposition under weaker assumptions},
  journal =      {Biometrics},
  year =         {2012},
  volume =       {68},
  pages =        {1019-1027},
}

@ARTICLE{VanderWeeleChiba2014,
  author =       {VanderWeele, T J and Chiba, Y},
  title =        {Sensitivity analysis for direct and indirect effects in the presence of exposure-induced mediator-outcome confounders},
  journal =      {Epidemiology, biostatistics, and public health},
  year =         {2014},
  volume =       {11},
}

@ARTICLE{VansteelandtBekaertLange2012,
  author =       {Vansteelandt, S. and Bekaert, M. and Lange, T.},
  title =        {Imputation Strategies for the Estimation of Natural Direct and Indirect Effects},
  journal =      {Epidemiologic Methods},
  year =         {2012},
  volume =       {1},
  pages =        {129-158},
}

@incollection{Hong10,
year={2010},
booktitle={Proceedings of the American Statistical Association, Biometrics Section},
title={Ratio of mediator probability weighting for estimating natural direct and indirect effects},
publisher={Alexandria, VA: American Statistical Association},
author={Guanglei Hong},
pages={2401–2415},
}

@ARTICLE{Hu11b,
  AUTHOR =       {M. Huber},
  TITLE =        {Treatment evaluation in the presence of sample selection},
  JOURNAL =      {Econometric Reviews},
    volume = {33},
  pages = {869-905},
  year =         {2014},
}

@ARTICLE{Heckman79,
  author = {J.J. Heckman},
  title = {Sample Selection Bias as a Specification Error},
  journal = {Econometrica},
  year = {1979},
  volume = {47},
  pages = {153-161}
}

@ARTICLE{Rubin74,
	author    = "D B Rubin",
	title	  = "Estimating Causal Effects of Treatments in Randomized and Nonrandomized Studies",
	year      = 1974,
	journal   = "Journal of Educational Psychology",
	volume    = 66,
	pages     = "688-701"}

@book{LittleRubin87,
	author 	  = "R. Little and D. Rubin",
	publisher = "Wiley",
	address   = "New York",
	title     = "Statistical Analysis with Missing Data",
	year      = "1987"}

@ARTICLE{Gr74,
  AUTHOR =       "Reuben Gronau",
  TITLE =        "Wage comparisons-a selectivity bias",
  JOURNAL =      "Journal of Political Economy",
  YEAR =         "1974",
  volume =       "82",
  pages =        "1119-1143",
  month =        "Nov. - Dec."
  }

@ARTICLE{He76,
  AUTHOR =       "J. J. Heckman",
  TITLE =        "The common structure of statistical models of truncation, sample selection and limited dependent variables and a simple estimator for such models",
  JOURNAL =      "Annals of Economic and Social Measurement",
  YEAR =         "1976",
  volume =       "5",
  pages =        "475-492"
  }

@ARTICLE{Daneve03,
  AUTHOR =       "M. Das and W. K. Newey and F. Vella",
  TITLE =        "Nonparametric Estimation of Sample Selection Models",
  JOURNAL =      "Review of Economic Studies",
  YEAR =         "2003",
  volume =       "70",
  pages =        "33-58"
}

@ARTICLE{AnBeKr04,
  AUTHOR =       {Joshua Angrist and Eric Bettinger and Michael Kremer},
  TITLE =        {Long-Term Educational Consequences of Secondary School Vouchers: Evidence from Administrative Records in Colombia},
  JOURNAL =      {American Economic Review},
  volume =       {96},
  pages =        {847-862},
  YEAR =         {2006},
}

@article{Ru76b,
     title = {Inference and Missing Data},
     author = {D B Rubin},
     journal = {Biometrika},
     volume = {63},
     pages = {581-592},
     year = {1976},
     }

@ARTICLE{BaKe86,
  AUTHOR =       {Reuben M Baron and David A Kenny},
  TITLE =        {The Moderator-Mediator Variable Distinction in Social Psychological Research: Conceptual, Strategic,
  and Statistical Considerations},
  JOURNAL =      {Journal of Personality and Social Psychology},
  YEAR =         {1986},
  volume =       {51},
  pages =        {1173-1182},
}

@article{Co57,
     title = {Analysis of Covariance: Its Nature and Uses},
     author = {Cochran, William G.},
     journal = {Biometrics},
      volume = {13},
      pages = {261-281},
     year = {1957},
        }

@INPROCEEDINGS{Pearl01,
  AUTHOR =       {J Pearl},
  TITLE =        {Direct and indirect effects},
  BOOKTITLE =    {Proceedings of the Seventeenth Conference on Uncertainty in Artificial Intelligence},
  YEAR =         {2001},
  pages =        {411-420},
  address =      {San Francisco},
  publisher =    {Morgan Kaufman},
}

@ARTICLE{ImKeYa10,
  AUTHOR =       {Kosuke Imai and Luke Keele and Teppei Yamamoto},
  TITLE =        {Identification, Inference and Sensitivity Analysis for Causal Mediation Effects},
  JOURNAL =      {Statistical Science},
  YEAR =         {2010},
  volume =       {25},
  pages =        {51-71},
}

@ARTICLE{PeSiva06,
  AUTHOR =       {M L Petersen and S E Sinisi and M J van der Laan},
  TITLE =        {Estimation of Direct Causal Effects},
  JOURNAL =      {Epidemiology},
  YEAR =         {2006},
  volume =       {17},
  pages =        {276-284},
}

@ARTICLE{RoGr92,
  AUTHOR =       {J M Robins and Sander Greenland},
  TITLE =        {Identifiability and Exchangeability for Direct and Indirect Effects},
  JOURNAL =      {Epidemiology},
  YEAR =         {1992},
  volume =       {3},
  pages =        {143-155},
}

@ARTICLE{JuKe81,
  AUTHOR =       {C M Judd and D A Kenny},
  TITLE =        {Process Analysis: Estimating Mediation in Treatment Evaluations},
  JOURNAL =      {Evaluation Review},
  YEAR =         {1981},
  volume =       {5},
  pages =        {602-619},
}

@ARTICLE{BaPe97,
  AUTHOR =       {Alexander Balke and Judea Pearl},
  TITLE =        {Bounds on Treatment Effects From Studies With Imperfect Compliance},
  JOURNAL =      {Journal of the American Statistical Association},
  YEAR =         {1997},
  volume =       {92},
  pages =        {1171-1176},
}

@ARTICLE{FlFl09,
  AUTHOR =       {Carlos A. Flores and A. Flores-Lagunes},
  TITLE =        {Identification and Estimation of Causal Mechanisms and Net Effects of a Treatment under Unconfoundedness},
  JOURNAL =      {IZA DP No. 4237},
  YEAR =         {2009},
}

@ARTICLE{FlFl10,
  AUTHOR =       {Carlos A. Flores and Alfonso Flores-Lagunes},
  TITLE =        {Nonparametric Partial Identification of Causal Net and Mechanism Average Treatment Effects},
  JOURNAL =      {mimeo, University of Florida},
  YEAR =         {2010},
}

@ARTICLE{Caietal08,
  AUTHOR =       {Z Cai and M Kuroki and J Pearl and J Tian},
  TITLE =        {Bounds on Direct Effects in the Presence of Confounded Intermediate Variables},
  JOURNAL =      {Biometrics},
  YEAR =         {2008},
  volume =       {64},
  pages =        {695-701},
}

@ARTICLE{Kaufmanetal05,
  AUTHOR =       {S Kaufman and J Kaufman and R MacLenose and S Greenland and C Poole},
  TITLE =        {Improved Estimation of Controlled Direct Effects in the Presence of Unmeasured Confounding of Intermediate Variables},
  JOURNAL =      {Statistics in Medicine},
  YEAR =         {2005},
  volume =       {24},
  pages =        {1683-1702},
}

@ARTICLE{Sj09,
  AUTHOR =       {Arvid Sj\"{o}lander},
  TITLE =        {Bounds on natural direct effects in the presence of confounded intermediate variables},
  JOURNAL =      {Statistics in Medicine},
  YEAR =         {2009},
  volume =       {28},
  pages =        {558-571},
}

@ARTICLE{TenHaveetal2007,
  AUTHOR =  {Ten Have, T. R. and Joffe, M. M. and Lynch, K. G. and Brown, G. K. and Maisto, S. A. and Beck, A. T. },
  TITLE =        {Causal mediation analyses with rank preserving models},
  JOURNAL =      {Biometrics},
  YEAR =         {2007},
  volume =       {63},
  pages =        {926-934},
}

@ARTICLE{Albert2008,
  AUTHOR =       {J M Albert},
  TITLE =        {Mediation analysis via potential outcomes models},
  JOURNAL =      {Statistics in Medicine},
  YEAR =         {2008},
  volume =       {27},
  pages =        {1282-1304},
}

@ARTICLE{VanderWeele2010,
  author =       {VanderWeele, T J},
  title =        {Bias formulas for sensitivity analysis for direct and indirect effects},
  journal =      {Epidemiology},
  year =         {2010},
  volume =       {21},
  pages =        {540-551},
}

@ARTICLE{AlbertNelson2011,
  AUTHOR =       {J. M. Albert and S. Nelson},
  TITLE =        {Generalized causal mediation analysis},
  JOURNAL =      {Biometrics},
  YEAR =         {2011},
  volume =       {67},
  pages =        {1028-1038},
}

@ARTICLE{TchetgenTchetgenShpitser2011,
  AUTHOR =       {Tchetgen Tchetgen, E. J. and Shpitser, I.},
  TITLE =        {Semiparametric theory for causal mediation analysis: Efficiency bounds, multiple robustness, and sensitivity analysis},
  JOURNAL =      {The Annals of Statistics},
  volume =       {40},
  pages =        {1816-1845},
  YEAR =         {2012},
}

@INPROCEEDINGS{Robins2003,
  AUTHOR =       {J M Robins},
  TITLE =        {Semantics of causal DAG models and the identification of direct and indirect effects},
  BOOKTITLE =    {In Highly Structured Stochastic Systems},
  YEAR =         {2003},
  editor =       {P.J. Green and N.L. Hjort and S. Richardson},
  pages =        {70-81},
  address =      {Oxford},
  publisher =    {Oxford University Press},
}

@ARTICLE{VanderWeele09,
  AUTHOR =       {Tyler J. VanderWeele},
  TITLE =        {Marginal Structural Models for the Estimation of Direct and Indirect Effects},
  JOURNAL =      {Epidemiology},
  YEAR =         {2009},
  volume =       {20},
  pages =        {18-26},
}

@ARTICLE{HuberSolovyeva2017,
  author =       {Martin Huber and Anna Solovyeva},
  title =        {Direct and indirect effects under sample selection and outcome attrition},
  journal =      {SES Working Paper 496, University of Fribourg},
  year =         {2018},
}

@ARTICLE{Huber2015,
  author =       {M. Huber},
  title =        {Causal pitfalls in the decomposition of wage gaps},
  journal =      {Journal of Business and Economic Statistics},
  year =         {2015},
  pages = {179-191},
  volume = {33},
}

@ARTICLE{Keeleetal2015,
  author =       {Keele, Luke and Tingley, Dustin and Yamamoto, Teppei},
  title =        {Identifying mechanisms behind policy interventions via causal mediation analysis},
  journal =      {Journal of Policy Analysis and Management},
  year =         {2015},
  volume =       {34},
  pages =        {937-963},
}

@ARTICLE{HeckmanPintoSavelyev2013,
  author =       {James Heckman and Rodrigo Pinto and Peter Savelyev},
  title =        {Understanding the Mechanisms Through Which an Influential Early Childhood Program Boosted Adult Outcomes},
  journal =      {American Economic Review},
  year =         {2013},
  volume =       {103},
  pages =        {2052-2086},
}

@ARTICLE{ImYa2011,
  AUTHOR =       {Kosuke Imai and Teppei Yamamoto},
  TITLE =        {Identification and Sensitivity Analysis for Multiple Causal Mechanisms: Revisiting Evidence from Framing Experiments},
  JOURNAL =      {Political Analysis},
  volume = {21},
  pages = {141-171},
  YEAR =         {2013},
}

@article{HuberLechnerMellace2017,
author = {Huber, Martin and Lechner, Michael and Mellace, Giovanni},
title = {Why Do Tougher Caseworkers Increase Employment? The Role of Program Assignment as a Causal Mechanism},
journal = {The Review of Economics and Statistics},
volume = {99},
pages = {180-183},
year = {2017},
}

@incollection{Bertrand,
	author    = {Bertrand, Marianne},
	year={2011},
	title={{New Perspectives on Gender}},
	pages={1543-1590},
	booktitle = "Handbook of Labor Economics",
	bvolume   = "4",
	editor    = "O. Ashenfelter and D. Card",
	publisher = "Elsevier"
}

@article{AzmatPet14,
title = "Gender and the labor market: What have we learned from field and lab experiments?",
journal = "Labour Economics",
volume = "30",
pages = "32 - 40",
year = "2014",
doi = "https://doi.org/10.1016/j.labeco.2014.06.005",
author = "Ghazala Azmat and Barbara Petrongolo"
}

@misc{NLSY792000 ,
author = "{Bureau of Labor Statistics, U.S. Department of Labor}",
year = "2001",
title = "{National Longitudinal Survey of Youth 1979 cohort, 1979-2000 (rounds 1-19)}",
note = "{Produced and distributed by the Center for Human Resource Research, The Ohio State University. Columbus, OH}"
}

@article{honore2006bounds,
  title={Bounds on parameters in panel dynamic discrete choice models},
  author={Honor{\'e}, Bo and Tamer, Elie},
  journal={Econometrica},
  volume={74},
  number={3},
  pages={611--629},
  year={2006},
  publisher={Wiley Online Library}
}

@article{molinari2008partial,
  title={Partial identification of probability distributions with misclassified data},
  author={Molinari, Francesca},
  journal={Journal of Econometrics},
  volume={144},
  number={1},
  pages={81--117},
  year={2008},
  publisher={Elsevier}
}

@article{freyberger2015identification,
  title={Identification and shape restrictions in nonparametric instrumental variables estimation},
  author={Freyberger, Joachim and Horowitz, Joel L},
  journal={Journal of Econometrics},
  volume={189},
  number={1},
  pages={41--53},
  year={2015},
  publisher={Elsevier}
}

@article{laffers2019bounding,
  title={Bounding average treatment effects using linear programming},
  author={Laff{\'e}rs, Luk{\'a}{\v{s}}},
  journal={Empirical Economics},
  volume={57},
  number={3},
  pages={727--767},
  year={2019},
  publisher={Springer}
}

@article{manski2007partial,
  title={Partial identification of counterfactual choice probabilities},
  author={Manski, Charles F},
  journal={International Economic Review},
  volume={48},
  number={4},
  pages={1393--1410},
  year={2007},
  publisher={Wiley Online Library}
}

@article{laffers2017sensitivity,
  title={Sensitivity of the bounds on the ATE in the presence of sample selection},
  author={Laff{\'e}rs, Luk{\'a}{\v{s}} and
   Nedela, Roman},
  journal={Economics Letters},
  volume={158},
  pages={84--87},
  year={2017},
  publisher={Elsevier}
}

@article{demuynck2015bounding,
  title={Bounding average treatment effects: A linear programming approach},
  author={Demuynck, Thomas},
  journal={Economics Letters},
  volume={137},
  pages={75--77},
  year={2015},
  publisher={Elsevier}
}

@article{romano2010inference,
  title={Inference for the identified set in partially identified econometric models},
  author={Romano, Joseph P and Shaikh, Azeem M},
  journal={Econometrica},
  volume={78},
  number={1},
  pages={169--211},
  year={2010},
  publisher={Wiley Online Library}
}

\clearpage
\newpage

\begin{appendices}

\section{Deriving bounds based on linear programming}\label{bounds}

\subsection{Optimization problems for bounds on mean potential outcomes}

{\footnotesize
  For $i = 1, \dots, n$, $K \in \{1, 2, 3\}$ and for a particular relaxation type $R \in \{ X1, X2, X3, M1, M2, M3, probit \}$ we denote
\begin{packed_item}
\item $\omega_i^{AK} = 1/q_i^{AK}$
\item $\bar{\omega}_i^{AK} = 1/(1-q_i^{AK})$,
\item $\epsilon_{i,R}^{AK} = \frac{|\hat p_{i,R}^{AK} - \hat p_i^{AK}|}{ \sqrt{\hat p_i^{AK}(1-\hat p_i^{AK})}}$
\item  $\epsilon^{AK,1}_{R} = \sum_{i=1}^n \frac{D_i \cdot S_i  \cdot \epsilon_{i,R}^{AK}}{\sum_{i=1}^n D_i \cdot S_i }$
\item  $\epsilon^{AK,0}_{R} = \sum_{i=1}^n \frac{(1-D_i) \cdot S_i  \cdot \epsilon_{i,R}^{AK}}{\sum_{i=1}^n (1-D_i) \cdot S_i }$
\end{packed_item}

}

\subsection{\texorpdfstring{ Bounds on $\hat E[Y(1,M(1))]$ }{TEXT}   }

{\footnotesize
\begin{eqnarray*}
\underset{ \omega^{A1},\omega^{A3}}{\min/\max} && \frac{1}{c_{11}} \cdot \sum_{i = 1}^{n} Y_i \cdot D_i \cdot S_i \cdot \omega_i^{A1} \cdot \omega_i^{A3}\\
&s.t.& \notag\\
\sum_{i=1}^n D_i \cdot S_i \cdot \omega_i^{A1} &=&  \sum_{i=1}^n \frac{D_i \cdot S_i}{\hat p_i^{A1}} , \notag\\
 \sum_{i=1}^n  D_i\cdot S_i \cdot \omega_i^{A3} &=& \sum_{i=1}^n \frac{D_i\cdot S_i } {\hat p_i^{A3}} ,   \notag\\
\sum_{i=1}^n  D_i \cdot S_i \cdot \omega_i^{A1} \cdot \omega_i^{A3}&=&  \sum_{i=1}^n \frac{D_i}{\hat p_i^{A1}} \frac{S_i}{\hat p_i^{A3}} = c_{11}   ,   \notag\\
\forall i:  \omega_i^{A1} &\leq& 1/ \left(\hat p_i^{A1} - \epsilon^{A1,1}_{R} \sqrt{\hat p_i^{A1}(1-\hat p_i^{A1})} \right), \notag\\
\omega_i^{A1} &\geq& 1/ \left(\hat p_i^{A1} + \epsilon^{A1,1}_{R} \sqrt{\hat p_i^{A1}(1-\hat p_i^{A1})} \right), \notag\\
\omega_i^{A3} &\leq& 1/ \left(\hat p_i^{A3} - \epsilon^{A3,1}_{R} \sqrt{\hat p_i^{A3}(1-\hat p_i^{A3})} \right), \notag\\
\omega_i^{A3} &\geq& 1/ \left(\hat p_i^{A3} + \epsilon^{A3,1}_{R} \sqrt{\hat p_i^{A3}(1-\hat p_i^{A3})} \right), \notag\\
\omega_i^{A1} &\geq& 0,  \notag\\
\omega_i^{A3} &\geq& 0.   \notag
\end{eqnarray*}
}


\subsection{\texorpdfstring{Bounds on $\hat E[Y(0,M(0))]$ }{TEXT}   }

{\footnotesize
\begin{eqnarray*}
\underset{ \omega^{A1},\omega^{A3}}{\min/\max} &&  \frac{1}{c_{00}} \cdot \sum_{i = 1}^{n} Y_i \cdot (1-D_i) \cdot S_i \cdot \omega_i^{A1} \cdot \omega_i^{A3}\\
&s.t.& \notag\\
\sum_{i=1}^n (1-D_i) \cdot S_i \cdot \omega_i^{A1} &=&  \sum_{i=1}^n \frac{(1-D_i) \cdot S_i}{\hat p_i^{A1}} , \notag\\
 \sum_{i=1}^n  (1-D_i)\cdot S_i \cdot \omega_i^{A3} &=& \sum_{i=1}^n \frac{(1-D_i)\cdot S_i } {\hat p_i^{A3}} ,   \notag\\
\sum_{i=1}^n  (1-D_i) \cdot S_i \cdot \omega_i^{A1} \cdot \omega_i^{A3}&=&  \sum_{i=1}^n \frac{(1-D_i)}{\hat p_i^{A1}} \frac{S_i}{\hat p_i^{A3}} = c_{00}   ,   \notag\\
\forall i:  \omega_i^{A1} &\leq& 1/ \left(\hat p_i^{A1} - \epsilon^{A1,0}_{R} \sqrt{\hat p_i^{A1}(1-\hat p_i^{A1})} \right), \notag\\
\omega_i^{A1} &\geq& 1/ \left(\hat p_i^{A1} + \epsilon^{A1,0}_{R} \sqrt{\hat p_i^{A1}(1-\hat p_i^{A1})} \right), \notag\\
\omega_i^{A3} &\leq& 1/ \left(\hat p_i^{A3} - \epsilon^{A3,0}_{R} \sqrt{\hat p_i^{A3}(1-\hat p_i^{A3})} \right), \notag\\
\omega_i^{A3} &\geq& 1/ \left(\hat p_i^{A3} + \epsilon^{A3,0}_{R} \sqrt{\hat p_i^{A3}(1-\hat p_i^{A3})} \right), \notag\\
\omega_i^{A1} &\geq& 0,  \notag\\
\omega_i^{A3} &\geq& 0.   \notag
\end{eqnarray*}
}



\subsection{\texorpdfstring{Bounds on  $\hat E[Y(1,M(0))]$ }{TEXT}   }

{\footnotesize
\begin{eqnarray*}
\underset{ \bar{\omega}^{A1}, \ \omega^{A2}, \ \omega^{A3}}{\min/\max} &&  \frac{1}{c_{10}} \cdot \sum_{i = 1}^{n} Y_i \cdot D_i \cdot S_i \cdot \bar{\omega}_i^{A1} \cdot (\omega^{A2}-1) \cdot \omega_i^{A3}\\
&s.t.& \notag\\
\sum_{i=1}^n D_i \cdot S_i \cdot \bar{\omega}_i^{A1} &=&  \sum_{i=1}^n \frac{D_i \cdot S_i}{\hat p_i^{A1}} , \notag\\
\sum_{i=1}^n D_i \cdot S_i \cdot \omega_i^{A2} &=&  \sum_{i=1}^n \frac{D_i \cdot S_i}{\hat p_i^{A1}} , \notag\\
 \sum_{i=1}^n  D_i\cdot S_i \cdot \omega_i^{A3} &=& \sum_{i=1}^n \frac{D_i\cdot S_i } {\hat p_i^{A2}} ,   \notag\\
\sum_{i=1}^n  D_i \cdot S_i \cdot \bar{\omega}_i^{A1}\cdot (\omega_i^{A2}-1)  \cdot \omega_i^{A3}&=&  \sum_{i=1}^n \frac{D_i \cdot S_i}{\hat p_i^{A1}\hat p_i^{A3}} \left(\frac{1}{\hat p_i^{A2}}-1 \right) \notag\\
\forall i:  \bar{\omega}_i^{A1} &\leq& 1/ \left(1-\hat p_i^{A1} - \epsilon^{A1,1}_{R} \sqrt{\hat p_i^{A1}(1-\hat p_i^{A1})} \right), \notag\\
\bar{\omega}_i^{A1} &\geq& 1/ \left(1-\hat p_i^{A1} + \epsilon^{A1,1}_{R} \sqrt{\hat p_i^{A1}(1-\hat p_i^{A1})} \right), \notag\\
\omega_i^{A2} &\leq& 1/ \left(\hat p_i^{A2} - \epsilon^{A2,1}_{R} \sqrt{\hat p_i^{A2}(1-\hat p_i^{A2})} \right), \notag\\
\omega_i^{A2} &\geq& 1/ \left(\hat p_i^{A2} + \epsilon^{A2,1}_{R} \sqrt{\hat p_i^{A2}(1-\hat p_i^{A2})} \right), \notag\\
\omega_i^{A3} &\leq& 1/ \left(\hat p_i^{A3} - \epsilon^{A3,1}_{R}\sqrt{\hat p_i^{A3}(1-\hat p_i^{A3})} \right), \notag\\
\omega_i^{A3} &\geq& 1/ \left(\hat p_i^{A3} + \epsilon^{A3,1}_{R} \sqrt{\hat p_i^{A3}(1-\hat p_i^{A3})} \right), \notag\\
\bar{\omega}_i^{A1} &\geq& 0,  \notag\\
\omega_i^{A2} &\geq& 0.   \notag\\
\omega_i^{A3} &\geq& 0.   \notag
\end{eqnarray*}
}



\subsection{\texorpdfstring{Bounds on  $\hat E[Y(0,M(1))]$ }{TEXT}   }

{\footnotesize
\begin{eqnarray*}
\underset{ \bar{\omega}^{A1}, \ \omega^{A2}, \ \omega^{A3}}{\min/\max} &&  \frac{1}{c_{01}} \cdot \sum_{i = 1}^{n} Y_i \cdot (1-D_i) \cdot S_i \cdot \omega_i^{A1} \cdot (\bar{\omega}^{A2}-1) \cdot \omega_i^{A3}\\
&s.t.& \notag\\
\sum_{i=1}^n (1-D_i) \cdot S_i \cdot \bar{\omega}_i^{A1} &=&  \sum_{i=1}^n \frac{(1-D_i) \cdot S_i}{\hat p_i^{A1}} , \notag\\
\sum_{i=1}^n (1-D_i) \cdot S_i \cdot \omega_i^{A2} &=&  \sum_{i=1}^n \frac{(1-D_i) \cdot S_i}{\hat p_i^{A1}} , \notag\\
 \sum_{i=1}^n  (1-D_i)\cdot S_i \cdot \omega_i^{A3} &=& \sum_{i=1}^n \frac{(1-D_i)\cdot S_i } {\hat p_i^{A3}} ,   \notag\\
\sum_{i=1}^n (1- D_i) \cdot S_i \cdot \omega_i^{A1}\cdot (\bar{\omega}_i^{A2}-1)  \cdot \omega_i^{A3}&=&  \sum_{i=1}^n \frac{(1-D_i) \cdot S_i}{\hat p_i^{A1}\hat p_i^{A3}}
\left(\frac{1}{1-\hat p_i^{A2}}-1 \right)  = c_{01},   \notag\\
\forall i:  \omega_i^{A1} &\leq& 1/ \left(\hat p_i^{A1} - \epsilon^{A1,0} \sqrt{1-\hat p_i^{A1}(\hat p_i^{A1})} \right), \notag\\
\omega_i^{A1} &\geq& 1/ \left(\hat p_i^{A1} + \epsilon^{A1,0} \sqrt{\hat p_i^{A1}(1-\hat p_i^{A1})} \right), \notag\\
\bar{\omega}_i^{A2} &\leq& 1/ \left(1-\hat p_i^{A2} - \epsilon^{A2,0} \sqrt{\hat p_i^{A2}(1-\hat p_i^{A2})} \right), \notag\\
\bar{\omega}_i^{A2} &\geq& 1/ \left(1-\hat p_i^{A2} + \epsilon^{A2,0} \sqrt{\hat p_i^{A2}(1-\hat p_i^{A2})} \right), \notag\\
\omega_i^{A3} &\leq& 1/ \left(\hat p_i^{A3} - \epsilon^{A3,0} \sqrt{\hat p_i^{A3}(1-\hat p_i^{A3})} \right), \notag\\
\omega_i^{A3} &\geq& 1/ \left(\hat p_i^{A3} + \epsilon^{A3,0} \sqrt{\hat p_i^{A3}(1-\hat p_i^{A3})} \right), \notag\\
\omega_i^{A1} &\geq& 0.   \notag\\
\bar{\omega}_i^{A2} &\geq& 0,  \notag\\
\omega_i^{A3} &\geq& 0.   \notag
\end{eqnarray*}
}

\pagebreak

\section{Descriptive statistics of the application}\label{app:app}

{\footnotesize
\begin{center}
	\renewcommand\arraystretch{0.65}
	\begin{longtable}{lcccc}
		\caption{Summary statistics and mean differences by gender} \label{tab:descr}  \\
		\hline\hline
		Variables        & Male($D=1$) & Female($D=0$) & Difference & \textit{p}-value \\ \hline
		\endfirsthead
		\caption*{Table~\ref{tab:descr}  -- continued from previous page}  \\
		\hline
		Variables        & Male($D=1$) & Female($D=0$) & Difference & \textit{p}-value \\ \hline
		\endhead
		\hline \multicolumn{5}{r}{{Continued on next page}} \\
		\endfoot
		
		\hline
		\endlastfoot
		
		\multicolumn{5}{l}{\textit{Outcome $Y$ (non-logged, refers to selected population with $S=1$)}}             	\\							
		Hourly wage                                          & 19.370     & 14.164       & 5.206     & 0.000             \\
		\hline \noalign{\smallskip}		
		
		\multicolumn{5}{l}{\textit{Mediators $M$ (refer to 1998 unless otherwise is stated)}} \\
		Married                                              & 0.566      & 0.568        & -0.002    & 0.882             \\
		Years married total since 1979                                 & 6.430      & 7.537        & -1.107    & 0.000             \\
		Northeastern region                                  & 0.153      & 0.155        & -0.002    & 0.857             \\
		North Central region                                 & 0.242      & 0.237        & 0.005     & 0.602             \\
		West region                                          & 0.206      & 0.195        & 0.011     & 0.244             \\
		South region (ref.)                                  & 0.399      & 0.414        & -0.015    & 0.205             \\
		Years lived in current region since 1979             & 14.839     & 15.246       & -0.407    & 0.000             \\
		Resides in SMSA                                      & 0.811      & 0.816        & -0.005    & 0.584             \\
		Years lived in SMSA  since 1979                      & 13.488     & 14.201       & -0.713    & 0.000             \\
		Less than high school (ref.)                         & 0.129      & 0.101        & 0.028     & 0.000             \\
		High school graduate                                 & 0.459      & 0.416        & 0.043     & 0.000             \\
		Some college                                         & 0.208      & 0.271        & -0.063    & 0.000             \\
		College or more                                      & 0.204      & 0.213        & -0.009    & 0.413             \\
		First job before 1975                                & 0.065      & 0.046        & 0.019     & 0.001             \\
		First job in 1976--79                                 & 0.115      & 0.128        & -0.013    & 0.083             \\
		First job after 1979 (ref.)                          & 0.821      & 0.825        & -0.004    & 0.623             \\
		Numer of jobs ever had                               & 10.555     & 9.239        & 1.316     & 0.000             \\
		Tenure with current employer (wks.)                  & 276.056    & 212.662      & 63.394    & 0.000             \\
		Industry: Primary sector                                       & 0.227      & 0.078        & 0.149     & 0.000             \\
		Industry: Manufacturing (ref.)                                 & 0.140      & 0.053        & 0.087     & 0.000             \\
		Industry: Transport                                            & 0.115      & 0.048        & 0.067     & 0.000             \\
		Industry: Trade                                                & 0.134      & 0.142        & -0.008    & 0.322             \\
		Industry: Finance                                              & 0.040      & 0.064        & -0.024    & 0.000             \\
		Industry: Services (business, personnel, and entertain.)        & 0.121      & 0.124        & -0.003    & 0.768             \\
		Industry: Professional services                                & 0.113      & 0.297        & -0.184    & 0.000             \\
		Industry: Public administration                                & 0.054      & 0.052        & 0.002     & 0.751             \\
		Years worked in current industry since 1982                     & 3.555      & 2.622        & 0.933     & 0.000             \\
		Manager                                              & 0.234      & 0.258        & -0.024    & 0.022             \\
		Technical occupation (ref.)                          & 0.039      & 0.038        & 0.001     & 0.907             \\
		Occupation in sales                                  & 0.067      & 0.082        & -0.015    & 0.021             \\
		Clerical occupation                                  & 0.056      & 0.212        & -0.156    & 0.000             \\
		Occupation in service                                & 0.102      & 0.163        & -0.061    & 0.000             \\
		Farmer or laborer                                    & 0.276      & 0.042        & 0.234     & 0.000             \\
		Operator (machines, transport)                       & 0.170      & 0.063        & 0.107     & 0.000             \\
		Years worked in current occupation since 1982        & 2.180      & 1.727        & 0.453     & 0.000             \\
		Employment status: employed                          & 0.877      & 0.748        & 0.129     & 0.000             \\
		Number of years employed status since 1979           & 13.204     & 11.271       & 1.933     & 0.000             \\
		Employed full time                                   & 0.846      & 0.599        & 0.247     & 0.000             \\
		Share of full-time employment 1994-98                & 0.896      & 0.658        & 0.238     & 0.000             \\
		Total number of weeks worked since 1979              & 661.794    & 560.408      & 101.386   & 0.000             \\
		Total number of weeks unemployed since 1979          & 62.343     & 49.744       & 12.599    & 0.000             \\
		Total number of weeks  out of labor force since 1979 & 146.118    & 265.276      & -119.158  & 0.000             \\
		Bad health prevents from working                     & 0.045      & 0.055        & -0.010    & 0.071             \\
		Years not working due to bad health since 1979       & 0.326      & 0.557        & -0.231    & 0.000             \\ \hline							
		\noalign{\smallskip}	
		
		\multicolumn{5}{l}{\textit{Pre-treatment covariates $X$}}	\\		
		Hispanic (ref.)                                      & 0.193      & 0.186        & 0.007     & 0.488             \\
		Black                                                & 0.287      & 0.297        & -0.010    & 0.413             \\
		White                                                & 0.520      & 0.517        & 0.003     & 0.840             \\
		Born in the U.S.                                     & 0.935      & 0.939        & -0.004    & 0.544             \\
		No religion                                          & 0.045      & 0.034        & 0.011     & 0.031             \\
		Protestant                                           & 0.501      & 0.500        & 0.001     & 0.957             \\
		Catholic (ref.)                                      & 0.352      & 0.352        & 0.000     & 0.967             \\
		Other religion                                       & 0.096      & 0.112        & -0.016    & 0.036             \\
		Mother born in U.S.                                  & 0.884      & 0.896        & -0.012    & 0.102             \\
		Mother’s educ. \textless high school (ref.)          & 0.376      & 0.421        & -0.045    & 0.000             \\
		Mother’s educ. high school graduate                  & 0.393      & 0.369        & 0.024     & 0.048             \\
		Mother’s educ. some college                          & 0.094      & 0.091        & 0.003     & 0.616             \\
		Mother’s educ. college/more                          & 0.076      & 0.071        & 0.005     & 0.411             \\
		Father born in U.S.                                  & 0.878      & 0.884        & -0.006    & 0.410             \\
		Father’s educ. \textless high school (ref.)          & 0.351      & 0.366        & -0.015    & 0.201             \\
		Father’s educ. high school graduate                  & 0.291      & 0.297        & -0.006    & 0.560             \\
		Father’s educ. some college                          & 0.087      & 0.076        & 0.011     & 0.105             \\
		Father’s educ. college/more                          & 0.131      & 0.117        & 0.014     & 0.085             \\
		Order of birth                                       & 3.195      & 3.259        & -0.064    & 0.256             \\
		Age in 1979                                          & 17.501     & 17.611       & -0.110    & 0.047             \\ \hline			
		\noalign{\smallskip}	
		
		\textit{Selection indicator S}	&	&	&		&         \\	
		Worked 1,000 hrs or more past year         & 0.867     &  0.696        & 0.171     & 0.000      \\
		\hline	
		\noalign{\smallskip}						
		Number of observations                                  & 3,162       & 3,496         &   .        &     . \\
		\hline
	\end{longtable}
\end{center}
}

\end{appendices}

\end{document}